\documentclass[%
 preprint, %linenumbers,
superscriptaddress,
 amsmath,amssymb,
 aps, physrev,
]{revtex4-2}

\usepackage{graphicx}% Include figure files
\usepackage{rotating} % For sidewaysfigure environment
\usepackage{dcolumn}% Align table columns on decimal point
\usepackage[version=4]{mhchem} 
\usepackage[utf8]{inputenc}
\usepackage{textgreek}
\usepackage{chemfig}
\usepackage{rotating}
\usepackage{graphicx}
\usepackage{tabularx}
\usepackage{adjustbox}
\usepackage{longtable}
\usepackage{chemformula}
\usepackage{array}
\usepackage{booktabs}

\usepackage{bm}% bold math
\begin{document}

\preprint{APS/123-QED}

\title{\textbf{Structure, Composition, and High-Field Superconductivity in Metal-Rich $\mathrm{\eta}$-Carbide-Type Compounds} 
}% 

\author{Manuele Balestra}
\affiliation {Department of Quantum Matter Physics, University of Geneva, CH-1211 Geneva, Switzerland}
 
\author{KeYuan Ma}
\affiliation{Max Planck Institute for Chemical Physics of Solids, 01187 Dresden, Germany}

\author{Harald O. Jeschke}
\affiliation {Research Institute for Interdisciplinary Science, Okayama University, Okayama 700-8530, Japan}

\author{Fabian O. von Rohr}
\affiliation{Department of Quantum Matter Physics, University of Geneva, CH-1211 Geneva, Switzerland}

\begin{abstract}
$\mathrm{\eta}$-Carbide-type compounds have recently emerged as a diverse class of materials in the study of superconductivity. These phases contribute to a growing family of metal-rich quantum materials that exhibit unusual superconducting properties emerging from complex metallic bonding. Several members of the $\mathrm{\eta}$-carbide-type phases have been found to be bulk superconductors -- such as \ce{Nb4Rh2C_{1-$\delta$}}, \ce{Ta4Rh2C_{1-$\delta$}}, \ce{Ti4Ir2O_{1-$\delta$}}, and \ce{Ti4Co2O_{1-$\delta$}} -- with transition temperatures up to $T_{\rm c} \approx$ 10 K and upper critical fields as high as $\mu_0 H_{\rm c2}(0) \approx$ 30 T. Whereas the transition temperatures may fall within the range typical for intermetallic superconductors, the pronounced violation of the weak-coupling Pauli limit in many of these crystallographically high-symmetry materials is noteworthy. Here, we review recent progress on superconducting $\mathrm{\eta}$-carbide-type phases, emphasizing how crystal symmetry, synthetic challenges, transition-metal composition, and electronic structure govern their superconducting properties. Furthermore, we outline open questions and future directions, including the possible discovery of new $\mathrm{\eta}$-carbide-type materials.

\end{abstract}

\maketitle

%\tableofcontents
\newpage

\section{Introduction}

$\mathrm{\eta}$-Carbides have long been studied as secondary phases in steels, where their formation can significantly influence mechanical performance.\cite{jack1973invited} These compounds, typically with a general formula \ce{A6C} or \ce{A3B3C} (where A and B are transition metals such as Fe, Mo, or W), crystallize in a cubic structure. These compounds often precipitate during heat treatment.\cite{taylor1952new} In tool and high-speed steels, a fine dispersion of $\mathrm{\eta}$-carbides can enhance hardness and wear resistance through precipitation strengthening. However, when these phases form coarsely, particularly along grain boundaries, they may act as brittle defects that compromise the material properties.\cite{hirotsu1972crystal} The term $\mathrm{\eta}$-carbides for the two cubic compositions \ce{Fe3W3C} and \ce{Fe4W2C} has been coined by Shuzo Takeda in 1930.\cite{Takeda1930a,Takeda1930b,Takeda1931, takeda1936metallographic} The previous term was double carbides, used for example in 1921~\cite{Daeves1921} and 1933~\cite{Adelskoeld1933}. Despite a long history of these phases, their electronic properties have received comparatively little attention.

$\mathrm{\eta}$-Carbide-type compounds are characterized by a high metal-to-nonmetal ratio, with transition-metal atoms forming a three-dimensional networks and the light elements occupying interstitial sites.\cite{Nohara2024} These elements can either be C, N, O, or B. As a result, the chemical bonding of these compounds is dominated by metal-metal interactions in contrary to other carbides.\cite{weinberger2018review,schwarz1987band} Carbide superconductors are a large family of superconductors that span a wide range of structural motifs: from simple binary compounds to layered intercalation systems. Binary carbides such as NbC, TaC, and \ce{Mo2C} exhibit superconductivity that is well described within conventional BCS theory, with critical temperatures typically in the range of 10-12 K.\cite{giorgi1962effect,simon1997superconductivity,kobayashi2019superconductivity} Layered graphite intercalation compounds, including \ce{CaC6} and \ce{YbC6}, also become superconducting through charge transfer between metal atoms and graphite sheets, again within a largely understood framework.\cite{weller2005superconductivity, mazin2005intercalant} Slightly more complex carbides, such as the intercalated fullerenes introduce additional electronic instabilities, yet their superconducting behavior remains closely linked to specific structural motifs.\cite{ganin2008bulk,ganin2010polymorphism} In contrast, the $\mathrm{\eta}$-carbide framework is characterized by a dense, cubic, metal-rich structure in which light elements occupy partially filled interstitial sites or vacancies. In fact, this metal-rich bonding renders simple valence electron counting inadequate and places $\mathrm{\eta}$-carbide-type compounds closer to intermetallic compounds with itinerant d-electron states.\cite{stadelmaier1969metal} This configurational flexibility and dominance of metal–metal interactions distinguish $\mathrm{\eta}$-carbide-type compounds from more chemically rigid carbide systems. In this context, $\mathrm{\eta}$-carbide-type superconductors are best viewed as part of a broader class of metal-rich quantum materials, in which light-element occupancy may influence -- but does not uniquely dictate -- the emergence of superconductivity.\cite{Nohara2024}

Recent work has advanced the understanding of superconductivity in compounds adopting the $\mathrm{\eta}$-carbide structure type, revealing a growing number of materials with robust bulk superconductivity and unusually large upper critical fields. For example, \ce{Nb4Rh2C_{1-$\delta$}} shows a superconducting transition temperature of $T_{\rm c} = 9.8$ K and an upper critical field of $\mu_0 H_{c2}(0) = 28.5$ T, significantly exceeding the weak-coupling Pauli paramagnetic limit.\cite{ku1985effect,shi2024nonmonotonic,ma2021superconductivity} Such behavior is noteworthy for these cubic and centrosymmetric metal-rich compounds, and points towards unusual features of the high-field superconductivity. Notably, thermodynamic signatures consistent with a possible Fulde–Ferrell–Larkin–Ovchinnikov (FFLO) state have been reported in \ce{Ti4Ir2O}, highlighting the potential for realizing exotic superconducting phases even in these systems.\cite{hu2023thermodynamic} $\mathrm{\eta}$-Carbide-type compounds exist over wide ranges of chemical compositions and allow for a high degree of atomic substitutions.\cite{nowotny1972crystal} This inherent structural flexibility and tunability provides considerable possibilities for the targeted modification and consequently for control of their physical properties. The $\mathrm{\eta}$-carbide family represents a promising family of compounds for the potential design and discovery of new superconductors. In this review, we explore the growing body of work on $\mathrm{\eta}$-carbide-type superconductors, examining these structural characteristics, the chemical synthesis, and the electronic properties.

%-----------------------------------------------------------------------------------------------------------------------------------
\section{Structure and Known Compounds of the $\mathrm{\eta}$-Carbide-Type Family}

Compounds belonging to the $\mathrm{\eta}$-structure-type family are isostructural and crystallize in the cubic $Fd\overline{3}m$ space group, adopting the \ce{Fe3W3C}-type structure. Commonly, these phases are also referred to as E9$_3$ phases \cite{Nyman1978}. 

Within the $\mathrm{\eta}$-carbide structure type, two closely related variants are commonly distinguished, conventionally denoted as $\mathrm{\eta}_1$ and $\mathrm{\eta}_2$. These variants differ primarily in stoichiometry and atomic site occupation.\cite{Weil1997} The $\mathrm{\eta}_1$ variant corresponds to compositions of the form \ce{A3B3X}, whereas the $\mathrm{\eta}_2$ variant adopts compositions of the form \ce{A4B2X}. Both variants crystallize in the same cubic space group and share the same underlying metal framework, but differ in the distribution of the two transition-metal species over the Wyckoff sites.

\begin{figure}[h!]
\centering
\includegraphics[width=0.8\textwidth]{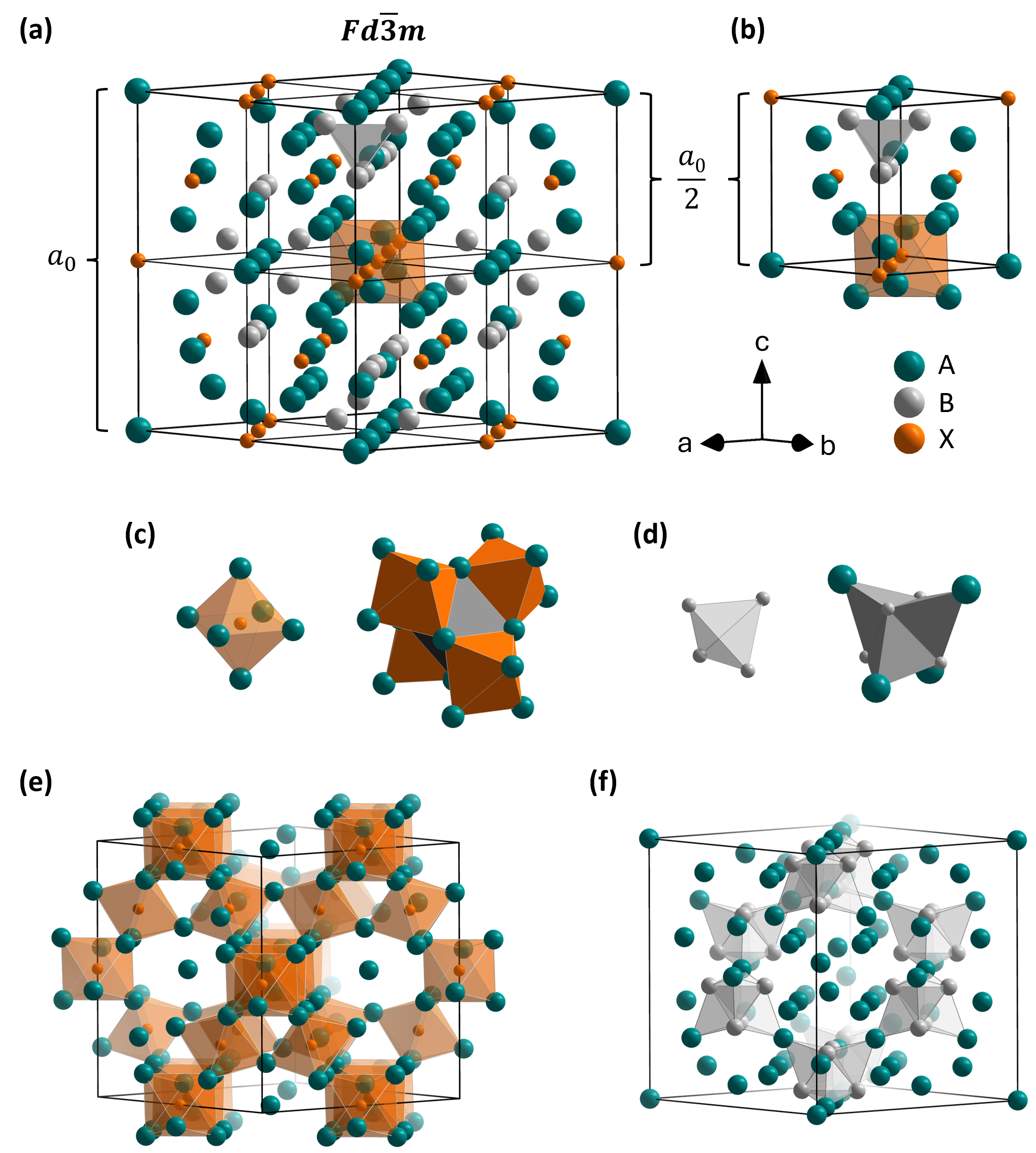}
\caption{Crystal structure of the $\mathrm{\eta}$-carbide–type compounds with space group $Fd\overline{3}m$. (a) Cubic unit cell with lattice parameter $a_{0}$, showing the positions of $A$ (teal), $B$ (gray), and $X$ (orange) atoms. (b) Half-unit-cell view highlighting the underlying metal framework. (c,d) Local coordination polyhedra of the X and B sites, respectively. (e) Network of corner-sharing $XA_{6}$ octahedra within the unit cell. (f) Corresponding arrangement of the metal sublattice, emphasizing the three-dimensional connectivity of the transition-metal network, formed by the \textit{stella quadrangula}.}
\label{fig:structure}
\end{figure}

The $\mathrm{\eta}$-carbide-type compounds can be understood as interstitially filled derivatives of the parent \ce{Ti2Ni}-type structure. \ce{Ti2Ni} crystallizes in the same distinct cubic structure type as the $\mathrm{\eta}$-carbide-type compounds, i.e. in space group $Fd\overline{3}m$. In the \ce{Ti2Ni} structure, nickel occupies the 32e and titanium the $48f$ Wyckoff sites. The \ce{Ti2Ni}-type structure contains 96 metal atoms per unit cell and can be described as comprising eight cubic subcells arranged in two alternating patterns, with interstitial regions between them \cite{Mueller1963}. An illustration of the unit cell and the aforementioned subunit cells is shown in Figs.~\ref{fig:structure}\,(a) and (b). 
A close structural relationship exists between the \ce{Ti2Ni} structure and both the \ce{Cr23C6} and $\mathrm{\eta}$-carbide structures, the latter sharing the same underlying metal matrix arrangement as \ce{Ti2Ni} \cite{Westgren1933}. A key feature of the \ce{Ti2Ni} and \ce{Cr23C6} structures is their ability to accommodate light non-metallic elements such as carbon, nitrogen, oxygen, or boron within interstitial sites, giving rise to the formation of a high variety of $\mathrm{\eta}$-carbide-type compounds \cite{taylor1952new,ma2021group,Souissi2018,Westgren1926,Waki2010,Waki2011,ku1985effect,Vandenberg1976}.
$\mathrm{\eta}$-carbide-type compounds contain 112 atoms in the unit cell \cite{ruan2022superconductivity,taylor1952new,Karlsson1951}.

In the $\mathrm{\eta}_1$ structure, the $A$ atoms occupy the $32e$ and $16d$ Wyckoff positions, while the $B$ atoms reside on the $48f$ sites. In contrast, the $\mathrm{\eta}_2$ structure is obtained by replacing the $A$ atoms on the $16d$ sites with $B$ atoms.\cite{Nyman1978,toth2014transition,Weil1997,Kuo1953} The light element $X$ occupies the interstitial $16c$ sites in both variants.\cite{Westgren1926} Additional derivatives of the $\mathrm{\eta}$ structure exist in which the interstitial species occupies alternative sites, such as the $8a$ positions, or a combination of $8a$ and $16c$ sites, further illustrating the structural flexibility of this family.\cite{bojarski1967neutron,Jeitschko1964} Unless stated otherwise, the following discussion refers to the $\mathrm{\eta}_2$ variant, which is most relevant for the superconducting compounds discussed here.

In these $\mathrm{\eta}$-carbide-type compounds, the 16 $A$ and 32 $B$ atoms form tetrahedra, and the 48 $B$ atoms form octahedra in which every second one is slightly distorted. The 16 $X$ atoms occupy the voids of the slightly distorted $XA_6$ octahedra (see, Figs.~\ref{fig:structure}\,(c)-(f)) \cite{Kuo1953}. The $XA_6$ octahedra are corner-sharing with each other.\cite{Weil1997,parthe1965neutron,Nohara2024} The $B$ atoms interpenetrate the network of these $XA_6$ octahedra, contributing to both structural stability and electronic properties \cite{Nohara2024}. Four $B$ atoms, together with four $A$ atoms, form a $B_4A_4$ cluster of four edge-sharing tetrahedra, also known as a \textit{stella quadrangula}, which are corner-sharing with each other and are shown in Fig.~\ref{fig:structure}\,(d) and (f) \cite{Weil1997,shi2025synergetic,Waki2011}. 
Studies on the formation tendencies of such $A_6X$ octahedra help predict the ordering of elements on specific crystallographic sites. Elements with the highest affinity for the interstitial $X$ position preferentially occupy the $A$ sites \cite{parthe1965neutron,Kuo1953}. 

An alternative description of the structure considers the $XA_6$ octahedra, formed by $A$ atoms around $X$ atoms, as corner-sharing units, and the four edge-sharing tetrahedra, composed of four $A$ and four $B$ atoms, as forming a \textit{stella quadrangula} network. These two sublattices interpenetrate to form a diamond cubic lattice, further contributing to the stabilization of the structure-type. \cite{Weil1997,Toth1971,shi2025synergetic,Waki2011}.

The $\mathrm{\eta}$-carbide structure exhibits a pronounced tolerance toward variations in light-element occupancy. Many $\mathrm{\eta}$-carbide-type compounds accommodate compositions of the form X$_{1-\delta}$ without changes to crystallographic symmetry or the underlying metal framework.\cite{Kuo1953,ma2021superconductivity} In some systems, incorporation of light elements such as C, N, O, or B is essential for stabilizing the $\mathrm{\eta}$-carbide framework, whereas in others the structure remains stable over a wide range of interstitial occupancies, even if the void positions are not occupied.\cite{ma2021group,Vandenberg1976,nevitt1960,Jeitschko1964} 

The stability of the \ce{Ti2Ni}-type metal framework correlates strongly with the atomic size ratio of the constituent metal species.\cite{nevitt1960,
rudman1967phase} Deviations of the radius ratio $R_{\rm A}$/$R_{\rm B}$ by more than approximately 8-10\% from an average value near 1.17 tend to destabilize the $\mathrm{\eta}$-carbide-type phase.\cite{nevitt1960,rudman1967phase, Kuo1953}

\section{\label{sec:level1}Synthesis Methods}

\begin{figure}[h!]
\centering
\includegraphics[width=0.8\textwidth]{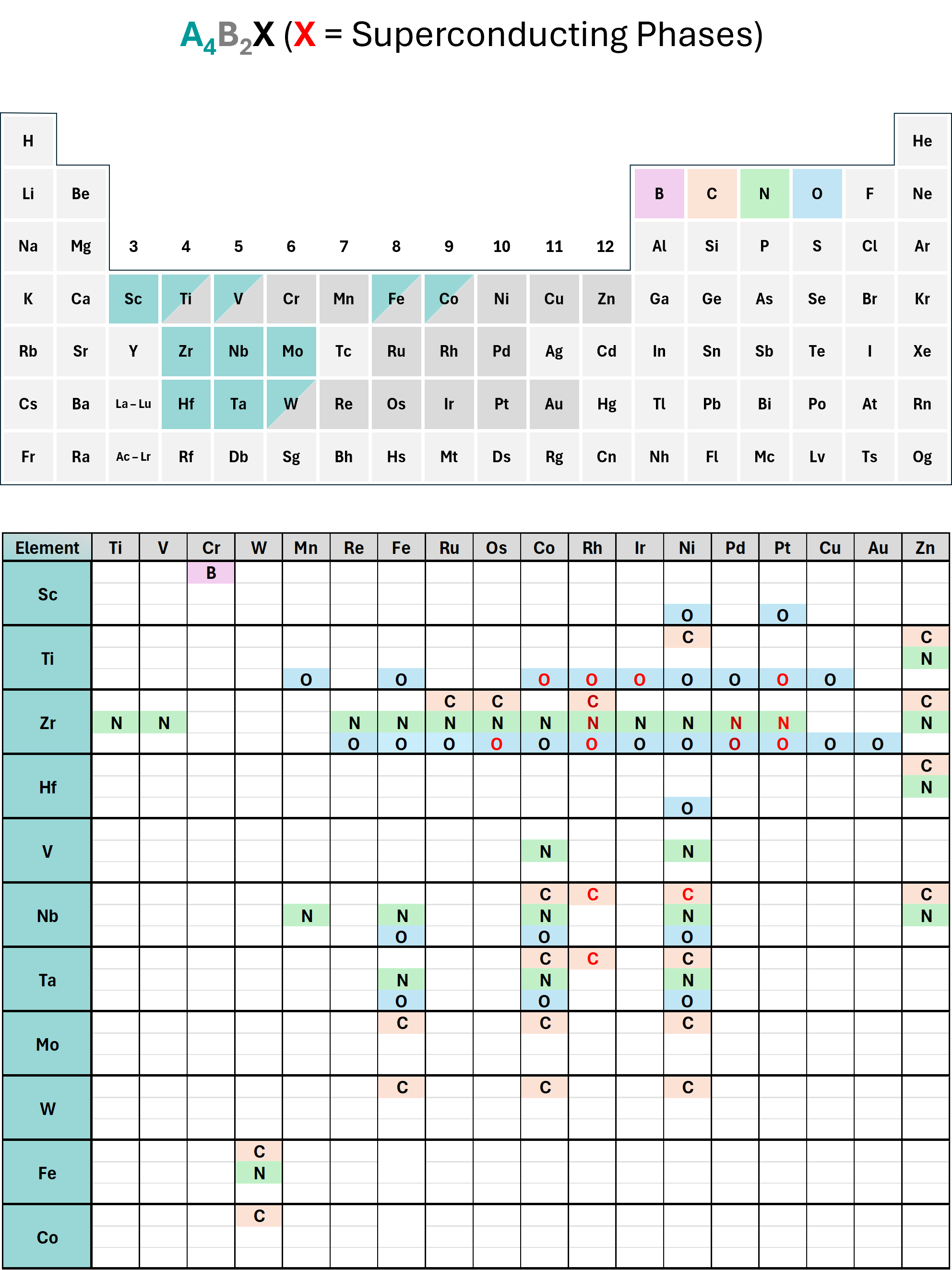}
\caption{Overview of known $\mathrm{\eta}$-carbide–type compounds with composition \ce{A4B2X}, organized by the transition-metal elements occupying the A and B sites of the cubic $Fd\overline{3}m$ framework. The incorporated light element X (C, N, or O) is indicated by color. Compositions reported to exhibit superconductivity are marked in red. The figure highlights both the wide chemical phase space accessible to the $\mathrm{\eta}$-carbide structure and the restricted region in which superconductivity has been observed.}
\label{fig:table_periodic_table}
\end{figure}

The synthesis of the $\mathrm{\eta}$-carbide-type compounds analyzed in this review relied predominantly on multi-step solid-state synthesis methods. Typically, the starting materials, whether elemental or as binary precursors, are mixed to ensure homogeneity and pressed into pellets. These pellets are then either sealed in a suitable reaction vessel, such as quartz, tantalum, niobium, or ceramics, or directly melted in an arc furnace under an inert argon atmosphere. The reactions can be carried out under vacuum or under a partial argon atmosphere, typically around 300 mbar. Commonly, the prereacted samples are post-annealed at temperatures between 800--1800 °C under vacuum or under a partial argon atmosphere for a prolonged time that can span up to 30 days. These annealing processes may be performed directly on ingots obtained from arc melting or on reground and repressed samples produced either from arc-melted ingots or from initially pressed powder precursors. For a more detailed insight into the different methods used to synthesize $\mathrm{\eta}$-carbide-type compounds, we refer to Table \ref{table:compounds_and_synthesis_1} and Table \ref{table:compounds_and_synthesis_2}, where we listed many of the published synthesis methods and conditions that were reported. We would like to highlight the contributions of Holleck \textit{et al.}, H. C. Ku and Nevitt \textit{et al.}, who have reported a range of compounds within the $\mathrm{\eta}$-compound family. However, due to missing details regarding the synthesis procedures used, these compounds were not included in these synthesis tables.\cite{ku1984new,Holleck1967, Holleck1967nuc} The various compounds reported by Holleck \textit{et al.}, H. C. Ku and Nevitt \textit{et al.} are, however, included in Fig.~\ref{fig:table_periodic_table}, which provides a comprehensive overview of the known $\mathrm{\eta}_2$-carbide-type compounds to date.

\begin{sidewaystable}
\fontsize{9}{7}\selectfont
\setlength{\tabcolsep}{4pt}
\renewcommand{\arraystretch}{2.0} % Adjust line spacing here (1.3 = slightly more spacing)
\centering
\begin{tabular}{
>{\raggedright\arraybackslash}p{3cm} % Cell length in column A
>{\raggedright\arraybackslash}p{12cm} % Cell length in column B
@{\hspace{1cm}} % extra horizontal space after column B 
>{\centering\arraybackslash}p{2.5cm} % Cell length in column C
>{\centering\arraybackslash}p{2cm} % Cell length in column D
}
\textbf{Compound} & \textbf{Synthesis Condition} & \textbf{Product Type} & \textbf{References} \\
\hline
\addlinespace[6pt]

\ce{Fe4W2N} & Synthesized via a two-step organometallic synthesis method: Metal chloride complexation in acetonitrile, followed by ammonolysis to form the nitride. & Polycrystalline & \cite{Weil1997} \\

\ce{Hf4Ni2N} & Mixed powders and cold pressed into pellet. Arc melted multiple times. & Not stated & \cite{Kotyk1970} \\

\ce{Hf4Ni2O} & Sintered pressed powder pellets at 1000 °C in quartz ampoules under vacuum. Arc melted. & Not stated & \cite{Kotyk1970} \\

\ce{Mo4Co2C} & Mixed powders and cold pressed into pellet under 20 tons. Sintered in zirconia/Alumina crucibles under vacuum at 1500 - 1800 °C. & Polycrystalline & \cite{Kuo1953} \\

\ce{Mo4Fe2C} & Mixed powders and cold pressed into pellet under 20 tons. Sintered in Zirconia/Alumina crucibles under vacuum at 1500 - 1800 °C. & Polycrystalline & \cite{Kuo1953} \\

\ce{Mo4Ni2C} & Mixed powders and cold pressed into pellet under 20 tons. Sintered in Zirconia/Alumina crucibles under vacuum at 1500 - 1800 °C. & Polycrystalline & \cite{Kuo1953} \\

\ce{Nb_{4-x}Rh_{2+x}C_{y}} & Mixed powders and cold pressed into pellet. Sintered at 1000 - 1250 °C for 3 days and quenched in water. & Polycrystalline & \cite{ku1985effect} \\

\ce{Ta4Ni2N} & Mixed powders and cold pressed into pellet. Arc melted multiple times. & Not stated & \cite{Kotyk1970} \\

\ce{Ta4Ni2O} & Sintered pressed powder pellets at 1000 °C. Wrapped pre-sintered pellets in Ta-foil and annealed at 1000 °C for 48h followed by 1350 °C for 24h. & Not stated\ polycrystalline & \cite{Kotyk1970} \\

\ce{Ta4Rh2C_{1-$\delta$}} & Mixed powders and cold pressed into pellet. Arc melted multiple times. Reground and pressed into pellet. Wrapped in Ta foil and sealed in quartz ampoule under 300 mbar Ar. Annealed at 1200 °C for 4 days and quenched in water. & Polycrystalline & \cite{ma2025discovery} \\

\ce{Ti4Co2O} & Mixed powders and cold pressed into pellet. Arc melted multiple times. Sealed in quartz ampoule under 300 mbar Ar. Annealed at 1000 °C for 30 days. & Single Crystals & \cite{Ma2022} \\

\ce{Ti4Co2O} & Mixed powders and cold pressed into dense rod. Arc melted multiple times. Sealed in quartz ampoule under 300 mbar Ar. Annealed at 1000 °C for 30 days. & Single Crystals & \cite{shi2025synergetic} \\

\ce{Ti4Co2O} & Mixed powders and cold pressed into pellet. Arc melted multiple times. Sealed in quartz ampoule under 300 mbar Ar. Annealed at 900 °C for 14 days and quenched in water. & Polycrystalline &  \cite{ku1984new} \\

\ce{Ti4Cu2O} & Arc Melted. Sealed in Vycor tubes. Annealed at 900 °C for 3 days. & Polycrystalline & \cite{Mueller1963} \\

\ce{Ti4Ir2O} & Mixed powders and cold pressed into pellet. Arc melted multiple times. Sealed in quartz ampoule under 300 mbar Ar. Annealed at 1400 °C for 14 days. & Single Crystals & \cite{Ma2022} \\

\ce{Ti4Ir2O} & Mixed powders and cold pressed into pellet. Placed in corundum crucible and sealed in Nb tube. Sintered at 1800 °C for 20h under Ar atmosphere. & Polycrystalline & \cite{ruan2022superconductivity} \\

\end{tabular}
\caption{Synthesis conditions for selected compounds.}
\label{table:compounds_and_synthesis_1}
\end{sidewaystable}

\begin{sidewaystable}
\fontsize{9}{7}\selectfont
\setlength{\tabcolsep}{4pt}
\renewcommand{\arraystretch}{2.0} % Adjust line spacing here between rows
\centering
\begin{tabular}{
>{\raggedright\arraybackslash}p{3cm} % Cell length in column A
>{\raggedright\arraybackslash}p{13.5cm} % Cell length in column B - increase to prevent linebreaks
@{\hspace{1cm}} % extra horizontal space after column B
>{\centering\arraybackslash}p{2.5cm} % Cell length in column C
>{\centering\arraybackslash}p{2cm} % Cell length in column D
}
\textbf{Compound} & \textbf{Synthesis Condition} & \textbf{Product Type} & \textbf{References} \\
\hline
\addlinespace[6pt]

\ce{Ti4Ni2C} & Mixed powders and pressed into a tablet. High-pressure high-temperature synthesis in an octahedral anvil configuration under 6 GPa at 1573 K for 40 minutes. & Single Crystals & \cite{Liu2024} \\

\ce{Ti4Ni2O} & Arc Melted. Sealed in Vycor tubes. Annealed at 900 °C for 3 days. & Polycrystalline & \cite{Mueller1963} \\

\ce{Ti4Pd2O} & Mixed elements. Arc melted, crushed and remelted the sample 5 times under an Ar atmosphere. Annealed at 873 K under vacuum. & Polycrystalline & \cite{Cantrell2002} \\

\ce{Ti4Rh2O} & Mixed powders and cold pressed into pellet. Arc melted multiple times. Sealed in quartz ampoule under 300 mbar Ar. Annealed at 1200 °C for 21 days. & Single Crystals & \cite{Ma2022} \\

\ce{W4Co2C} & Mixed powders and cold pressed into pellet. Sintered at unspecified temperature in sealed quartz ampoules under vacuum. Arc melted. Sealed in quartz ampoules under vacuum. Annealed at 1100 °C for 75h followed by 200h at 1000 °C. & Polycrystalline & \cite{Pollock1970} \\

\ce{W4Fe2C} & Mixed powders and cold pressed into pellet. Sintered at unspecified temperature in sealed quartz ampoules under vacuum. Arc melted. Sealed in quartz ampoules under vacuum. Annealed at 1200 °C for 30h followed by 200h at 1000 °C. & polycrystalline & \cite{Pollock1970} \\

%\ce{Zr3V3O_{x}} & Mixed powders and cold pressed into pellet. Act melted. Annealed at 800 °C for 400h. & Polycrystalline & \cite{Zavaliy1999} \\

\ce{Zr4Fe2O_x} & Mixed powders and cold pressed into pellet. Arc melted. Annealed at 1000 °C for 100h. & Polycrystalline & \cite{Zavaliy1999} \\

\ce{Zr4Ni2O_x} & Mixed powders and cold pressed into pellet. Arc melted. Annealed as-cast samples at 1000 °C for 300h. & not specified & \cite{Lavrentyev2013-mb} \\

\ce{Zr4Ni2O} & Mixed Zr and Ni metals and arc melted. Ground to small pieces. Mixed with appropriate amount of \ce{NaClO4} and sealed in quartz ampoule. Heated to 600 °C for 4 days in tube furnace where \ce{NaCl} crystals formed at the cool end. Crushed the oxide, pressed into pellet and annealed at 1250 °C. & Polycrystalline & \cite{Mackay1994} \\

\ce{Zr4Pd2O} & Mixed elements and pressed into pellet. Arc melted under purified Ar gas. Annealed at 1273 K under high vacuum for 8 days. & Polycrystalline & \cite{Leonard1991} \\

\ce{Zr4Pd2O} & Mixed elements. Arc melted, crushed and remelted the sample 5 times under an Ar atmosphere. Annealed at 873 K under vacuum. & Polycrystalline & \cite{Cantrell2002} \\

\ce{Zr4Pt2O} & Mixed elements and cold pressed into pellet. Arc melted multiple times. & Polycrystalline & \cite{Leonard1991} \\

\ce{Zr4Rh2C_{1-$\delta$}} & Mixed powders without Zr and cold pressed into pellet. Arc melted the pellet together with the Zr plate and remelted multiple times. Reground and pressed into pellet. Sealed in quartz ampoule under vacuum. Annealed at 800 °C for 10 days. & Polycrystalline & \cite{watanabe2023observation} \\

\ce{Zr4Rh2O_{x} ($x \leq 0.7$)} & Mixed powders and cold pressed into pellet. Arc melted multiple times. Reground and pressed into pellet. Sealed in quartz ampoule under 300 mbar Ar. Annealed at 1000 °C for 10 days. & Polycrystalline & \cite{ma2019superconductivity} \\

\ce{Zr4Rh2O_{x} ($x \geq 0.7$)} & Mixed powders and cold pressed into pellet. Arc melted multiple times. Reground and pressed into pellet. Sealed in quartz ampoule under 300 mbar Ar. Annealed at 800 °C for 10 days. & Polycrystalline & \cite{ma2019superconductivity} \\
\end{tabular}
\caption{Synthesis conditions for selected compounds.}
\label{table:compounds_and_synthesis_2}
\end{sidewaystable}

\section{Stoichiometry and Composition of $\mathrm{\eta}$-Carbide-Type Superconductors}

Several decades ago, superconductivity was reported in a small number of $\mathrm{\eta}$-carbide–type compounds with general composition \ce{A4B2X} \cite{poole1999handbook,ku1984new,ku1985effect,RevModPhys.35.1}. These early studies primarily established the existence of superconducting transitions and provided approximate values of the critical temperature $T_{\rm c}$, but did not include systematic measurements of thermodynamic, magnetic, or transport properties. As a result, the superconducting state in these materials remained poorly characterized for many years.

In addition, the reported values of $T_{\rm c}$ often show significant scatter across different studies. This variability is likely related to the inherent difficulty of synthesizing phase-pure $\mathrm{\eta}$-carbide-type samples with well-controlled stoichiometry, particularly with respect to light-element occupancy and site disorder. A prime example of the challenges with early reports is found in the η-carbide–type oxide \ce{Zr4Rh2O_x}. Early literature listings cited a transition temperature near 11.8 K for a “Zr–Rh–O” phase attributed to the $\mathrm{\eta}$-carbide structure, but these reports lacked detailed characterization and stoichiometric control.\cite{RevModPhys.35.1} Systematic investigations of \ce{Zr4Rh2O_x} with controlled oxygen content reveal that the superconducting transition temperature is much lower and strongly dependent on O occupancy: bulk superconductivity is observed with $T_{\rm c} \approx$ 2.8 K for \ce{Zr4Rh2O_{0.7}} and $T_{\rm c} \approx$ 4.7 K for \ce{Zr4Rh2O}, based on resistivity and magnetization measurements on phase-pure $\mathrm{\eta}$-carbide-type samples.\cite{ma2019superconductivity} 

Another illustrative case is \ce{Zr3V3O}, which stood out as the only reported example of an $\mathrm{\eta}$-carbide-type compound with the 3--3--1 ($\mathrm{\eta}_1$) stoichiometry for which superconductivity was previously claimed.\cite{RevModPhys.35.1,rotella1983deuterium} These early reports attributed a superconducting transition of $T_{\rm c} \approx$ 7.5 K to this phase; however, subsequent investigations on phase-pure samples have failed to detect any superconducting transition above 1.8 K associated with this phase.\cite{Waki2012Zr3V3O,Ma2022} This discrepancy indicates that the originally reported superconductivity was likely associated with secondary phases and that it is not an intrinsic property of this $\mathrm{\eta}_1$-type \ce{Zr3V3O} compound. Henceforth, to date there is not a single $\mathrm{\eta}_1$-type superconductor known, to the best of our knowledge.

Another important aspect to consider is the role of the light element occupying the interstitial site of the $\mathrm{\eta}$-carbide framework. The influence of light-element occupancy on superconductivity has been demonstrated particularly clearly in the Ti--Co system. In this system, the oxygen-free compound \ce{Ti2Co} exhibits an onset to superconductivity below 0.7 K. The isostructural suboxide \ce{Ti4Co2O}, in which oxygen occupies the interstitial site of the $\mathrm{\eta}$-carbide-type structure, has been synthesized as a phase-pure compound and found to be a bulk type-II superconductor with a transition temperature of $T_{\rm c} \approx$ 2.7 K, accompanied by a normalized specific-heat jump $\Delta C/\gamma T_{\rm c}$ demonstrating the bulk superconductivity. This stark difference demonstrates that the incorporation of void-filling oxygen in $\mathrm{\eta}$-carbide–type frameworks has a critical effect on the electronic properties of these phases.\cite{ma2021group,shi2025synergetic}

\begin{figure}[h!]
\centering
\includegraphics[width=0.45\textwidth]{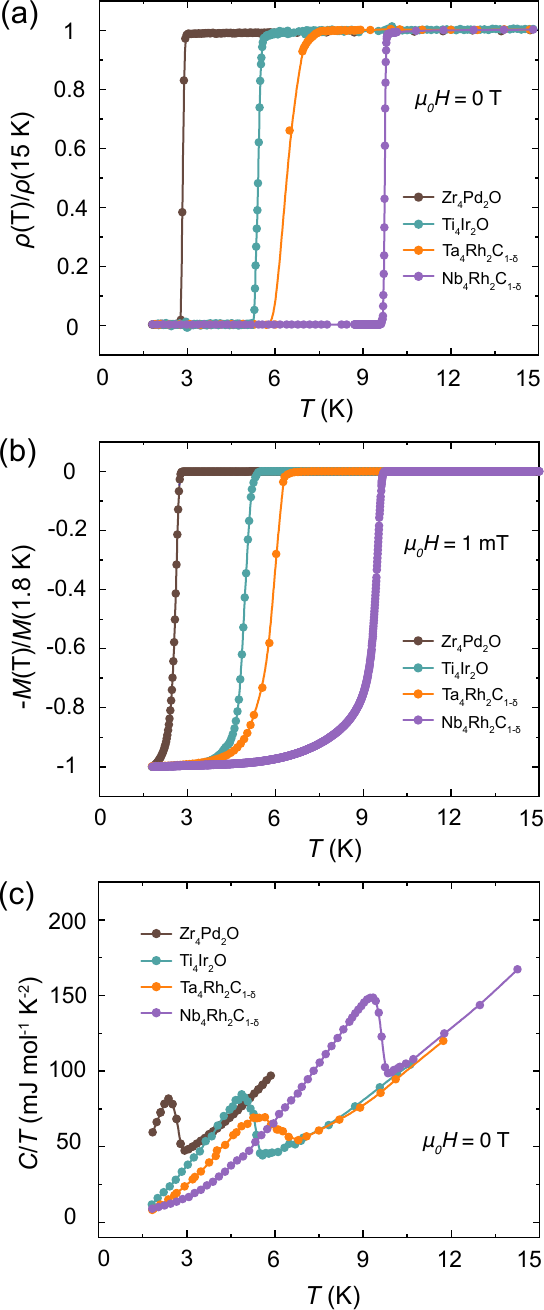}
\caption{Superconducting properties of Zr$_4$Pd$_2$O, Ti$_4$Ir$_2$O, Ta$_4$Rh$_2$C$_{1-\delta}$, and Nb$_4$Rh$_2$C$_{1-\delta}$. (a) normalized temperature-dependent resistivity under zero-field, (b) ZFC magnetization in an external field of 1 mT. (c)Temperature-dependent specific heat capacity  under zero-field in the vicinity of the superconducting transitions. Data taken and replotted from \cite{watanabe2023observation, ma2021superconductivity,ma2021group, ma2025discovery}.} 
\label{fig:supercon}
\end{figure}

\section{Establishing Bulk Superconductivity in $\mathrm{\eta}$-Carbide-Type Compounds}

In recent years, advances in synthesis have enabled the preparation of phase-pure $\mathrm{\eta}$-carbide–type superconductors, allowing their superconducting properties to be systematically characterized by temperature-dependent resistivity, magnetization, and specific-heat measurements. \cite{ma2019superconductivity,ma2021group,ruan2022superconductivity,watanabe2023observation} These studies have established reliable superconducting transition temperatures and, importantly, have confirmed the bulk nature of superconductivity in this class of materials. In addition to validating previously reported superconducting phases, this renewed effort has also led to the discovery of previously unknown $\mathrm{\eta}$-carbide-type superconductors, as in, e.g., \ce{Ta4Rh2C_{1-$\delta$}} or \ce{Zr4Pd2O}.\cite{ma2025discovery,watanabe2023observation}

Table~\ref{tab:table2} summarizes the currently known $\mathrm{\eta}$-carbide-type superconductors together with their superconducting transition temperatures $T_{\rm c}$ and zero-temperature upper critical fields $H_{\rm c2}(0)$. Reported values of $T_{\rm c}$ span the range from approximately 2.1 K to 9.8 K. Among the known compounds, \ce{Nb4Rh2C_{1-$\delta$}} exhibits the highest transition temperature at ambient pressure, with $T_{\rm c} = 9.8$ K.\cite{ma2021superconductivity} 

Recently, the physical properties of several $\mathrm{\eta}$-carbide–type superconductors—including \ce{Zr4Pd2O}, \ce{Ti4Ir2O}, \ce{Nb4Rh2C_{1-$\delta$}}, and related compounds—have been systematically characterized. Fig.~\ref{fig:supercon} summarizes representative temperature-dependent resistivity, magnetization, and specific-heat measurements for \ce{Zr4Pd2O}, \ce{Ti4Ir2O}, \ce{Ta4Rh2C_{1-$\delta$}}, and \ce{Nb4Rh2C_{1-$\delta$}} in the vicinity of their superconducting transitions at $\mu_{0}H = 0$ T.\cite{watanabe2023observation,ma2021group,ma2025discovery,ma2021superconductivity} In all cases, the resistivity drops sharply to zero, a pronounced diamagnetic response is observed under zero-field-cooled conditions, and clear anomalies appear in the specific heat at the superconducting transition temperatures. Together, these signatures unambiguously establish the presence of bulk superconductivity in these $\mathrm{\eta}$-carbide-type compounds.

Quantitative analysis of the specific-heat data further supports this conclusion. The normalized specific-heat jumps $\Delta C/\gamma T_{\rm c}$ for \ce{Zr4Pd2O}, \ce{Ti4Ir2O}, \ce{Ta4Rh2C_{1-$\delta$}}, and \ce{Nb4Rh2C_{1-$\delta$}} are found to be 1.58, 1.80, 1.56, and 1.64, respectively, exceeding the weak-coupling BCS value of 1.43. \cite{watanabe2023observation,ma2021group,ma2025discovery,ma2021superconductivity} These enhanced values are consistent with moderately strong electron–phonon coupling in these materials. Notably, the particularly large $\Delta C/\gamma T_{\rm c}$ value of 1.80 observed in \ce{Ti4Ir2O} may point to stronger coupling effects.

In the normal state, the resistivity of the reported $\mathrm{\eta}$-carbide-type superconductors decreases monotonically upon cooling, as expected for their metallic transport behavior. Temperature-dependent magnetization measurements above the critcial temperature reveal a weak, nearly temperature-independent Pauli-paramagnetic response from consistent with a predominantly itinerant-electron normal state.

\begin{table}[b]
\caption{\label{tab:table2}
Summary of all reported $\mathrm{\eta}$-carbide-type superconductors.}
\begin{ruledtabular}
\begin{tabular}{ccccc}
 Compounds\footnotemark[1] & $T_{\rm c}$ (K) &$H_{\rm c2}(0)$ (T) & $H_{\rm Pauli}$\footnotemark[2](T)& Reference \\
\hline
Nb$_4$Rh$_2$C & 8.5-9.8 & 28.5 & 18.2 & \cite{ku1984new,ku1985effect, ma2021superconductivity}\\
Ta$_4$Rh$_2$C & 6.4 & 17.4 & 11.9 & \cite{ma2025discovery}\\
Ti$_4$Ir$_2$O & 5.1-5.7& 16.5 & 9.9 & \cite{ma2021group,das2024ti,ruan2022superconductivity}\\
Zr$_4$Rh$_2$O & 2.7-4.7& 6.1 & 8.7 & \cite{ma2019superconductivity}\\
Zr$_4$Rh$_2$N & 3.0-3.1& - & - & \cite{ku1984new}\\
Zr$_4$Os$_2$O & 2.0-3.0& - & - & \cite{ku1984new}\\
Nb$_4$Ni$_2$C & 2.2-2.9 & - & - & \cite{ku1984new}\\
Ti$_4$Rh$_2$O & 2.8 & 5.2 & 5.2& \cite{ma2021group}\\
Ti$_4$Co$_2$O & 2.7 & 7.1 & 5.0& \cite{ma2021group}\\
Zr$_4$Pd$_2$O & 2.7 & 6.7 & 5.0& \cite{watanabe2023observation}\\
Zr$_4$Pd$_2$N & 1.3-2.1 & - & -& \cite{ku1984new}\\
Ti$_4$Pt$_2$O & 2.3-2.5 & - & -&\cite{ku1984new}\\
Zr$_4$Pt$_2$N & 2.1-2.3 & - & -& \cite{ku1984new}\\
Zr$_4$Pt$_2$O & 1.3-1.6 & - & -& \cite{ku1984new}\\

\end{tabular}
\end{ruledtabular}
\footnotetext[1]{We list the reported compounds in the ideal chemical stoichiometric ratio for clarity of comparison.}
\footnotetext[2]{Here, $H_{\rm Pauli}$ is calculated using $H_{\rm Pauli} \approx 1.86{\rm [T/K]} \cdot T_{\rm c}$  
}
\end{table}

\section{High upper critical field in $\mathrm{\eta}$-carbide-type superconductors}

A striking feature of several $\mathrm{\eta}$-carbide–type superconductors is their unusually large upper critical fields, which in some cases substantially exceed expectations based on weak-coupling superconductivity. The upper critical field $H_{\rm c2}$ marks the magnetic field at which superconductivity is suppressed. In general, superconductivity is destroyed by an applied magnetic field through two principal processes: orbital pair breaking and Pauli paramagnetic (i.e. the Zeeman effect) pair breaking.\cite{clogston1962upper,maki1966effect,kirshenbaum2013pressure}

The orbital-limiting effect causes Cooper pair breaking by inducing a momentum in the single-particle spectrum that exceeds the superconducting gap. The Pauli paramagnetic effect corresponds to the Zeeman splitting energy of electronic spin exceeding the superconducting gap energy, at which point it becomes energetically unfavorable for electrons to pair with opposite spins. The orbital-limiting effect is commonly the dominant pair-breaking effect close to the $T_{\rm c}$ of a superconductor, while at low temperatures, far away from the $T_{\rm c}$, the Pauli paramagnetic effect is dominant.\cite{tinkham2004introduction}

The Pauli paramagnetic effect imposes an upper limit on the critical field of spin-singlet superconductors, known as the Pauli paramagnetic limit, which can be expressed by the following equation \cite{altarawneh2012superconducting}:

\begin{equation}
    \mu_0 H_{\rm Pauli} = \frac{\sqrt{2}\Delta_0}{{g} \mu_{\rm B}} 
\end{equation}

where $\mu_{\rm B}$ is the Bohr magneton, $g$ is the effective $g$-factor, and $\Delta_0$ is the superconducting gap. For conventional weak-coupling BCS superconductors $\Delta_0$= 1.76$k_{\rm B}$$T_{\rm c}$, (where $k_{\rm B}$ is the Boltzmann constant) and  $g$=2 (free-electron $g$-factor), the Pauli limit is approximated as: 

\begin{equation}
    \mu_0 H_{\rm Pauli} = \frac{\Delta_0}{\sqrt{2} \mu_{\rm B}} \approx 1.86{\rm [T/K]} \cdot T_{\rm c}
\end{equation}

The conventional weak-coupling BCS Pauli paramagnetic limit is respected by the vast majority of type-II superconductors. In most cases, the experimentally observed upper critical fields remain well below this limit. Consequently, superconductors that exhibit exceptionally large upper critical fields approaching or exceeding the Pauli limit are widely regarded as anomalous and have long been considered a hallmark of unconventional or nontrivial superconducting behavior.

Recently, the systematical investigations on some $\mathrm{\eta}$-carbide-type superconductors have shown that the $\mathrm{\eta}$-carbide-type superconductors have very high upper critical fields, which are for some materials even higher than the weak-coupling Pauli limit. Notably, the Nb$_4$Rh$_2$C$_{1-\delta}$ superconductor with a $T_{\rm c}$ = 9.8 K shows a high upper critical field of $\mu_0 H_{c2}(0)$ = 28.5 T. 

\begin{figure}[h!]
\centering
\includegraphics[width=\linewidth]{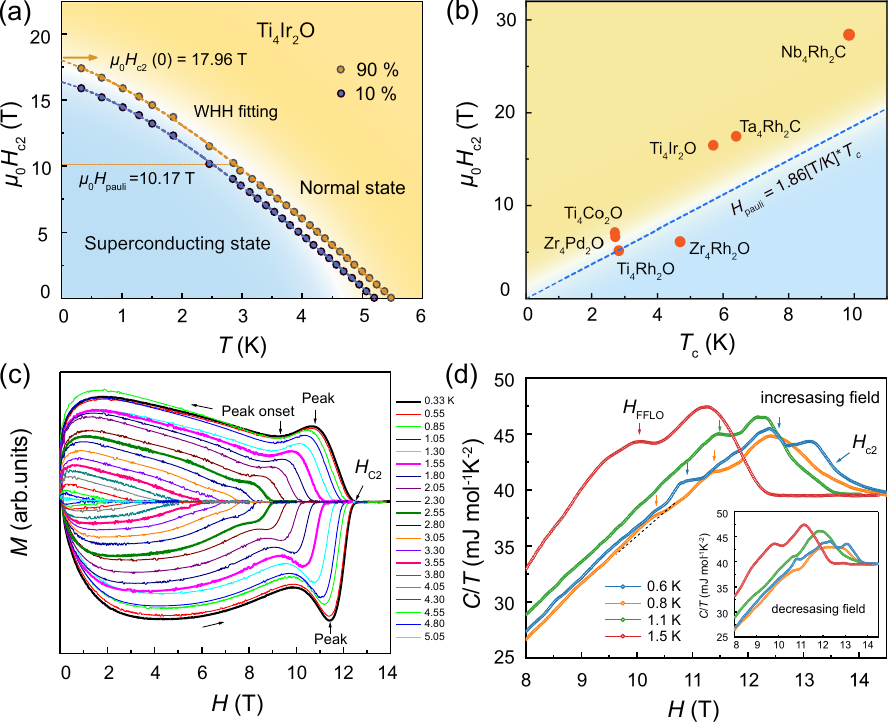}
\caption{Upper critical field behavior and high-field thermodynamic signatures in $\mathrm{\eta}$-carbide–type superconductors. (a) $\mu_{0}H_{\rm c2}(T)$ for \ce{Ti4Ir2O}, showing pronounced enhancement relative to the weak-coupling Pauli limit and comparison with WHH theory (reprinted and modified from Ref.\cite{das2024ti} licensed under CC BY 4.0.). (b) $\mu_{0}H_{\rm c2}(0)$ as a function of $T_{\rm c}$ for several $\mathrm{\eta}$-carbide superconductors; the dashed line indicates the Pauli paramagnetic limit. Figure (a) and (b): Reanalysis of data from Refs. \cite{das2024ti,ma2021superconductivity,ma2025discovery,ma2019superconductivity,ma2021group}. (c) Temperature-dependent magnetization $M(H)$ of \ce{Ti4Ir2O}, with the upper critical field identified from the field-induced features. (d) Field-dependent specific heat $C/T$ for \ce{Ti4Ir2O}, displaying anomalies near $H_{\rm c2}$ and additional structure at high fields, suggestive of an unconventional high-field superconducting phase. For (c) and (d) figures adapted and modified from \cite{hu2023thermodynamic} licensed under CC BY 4.0.}
\label{fig:highfield}
\end{figure}

Among the systematically measured $\mathrm{\eta}$-carbide-type superconductors, Ti$_4$Co$_2$O, Ti$_4$Ir$_2$O, Nb$_4$Rh$_2$C$_{1-\delta}$, Ta$_4$Rh$_2$C$_{1-\delta}$, and Zr$_4$Pd$_2$O were found to have $\mu_0 H_{\rm c2}{\rm (0)}$ larger than their weak-coupling Pauli limits.  Fig.~\ref{fig:highfield}\,(a) shows the temperature dependence of $\mu_0 H_{\rm c2}{\rm (0)}$ of Ti$_4$Ir$_2$O, while Fig.~\ref{fig:highfield}\,(b) shows the extracted upper critical field values $\mu_0 H_{\rm c2}{\rm (0)}$ for the different known $\mathrm{\eta}$-carbide superconductors versus their critical temperature $T_{\rm c}$.

Superconductivity beyond the weak-coupling BCS Pauli limit can be stabilized in the presence of finite-momentum pairing or strong spin-orbit coupling. The former leads to the Fulde–Ferrell–Larkin–Ovchinnikov (FFLO) state, in which Cooper pairs acquire a finite center-of-mass momentum and the upper critical field can exceed the Pauli limit by a limited amount. Strong spin-orbit coupling, on the other hand, can substantially weaken Pauli paramagnetic pair breaking, for example, through a renormalization of the effective electronic $g$-factor, thereby allowing the upper critical field to exceed the Pauli limit by a much larger margin \cite{norman1990magnetic,yoshimura2025g}.

The FFLO state is difficult to observe experimentally because it requires a series of strict conditions, like low temperature, high magnetic field, high quality crystal samples, and strong coupling.\cite{kinnunen2018fulde} Hu \textit{et al.} explored the high upper critical field of Ti$_4$Ir$_2$O using thermodynamic measurement probes and observed a characteristic upturn in the upper critical field line within the low-temperature, high-magnetic-field regime of the magnetic phase diagram.\cite{hu2023thermodynamic} The authors attributed these anomaly signatures to a possible formation of a Fulde-Ferrell-Larkin-Ovchinnikov (FFLO) state, which can provide a plausible explanation for the violation of the Pauli limit in this superconductor. Figs.~\ref{fig:highfield}\,(c) and (d) show the temperature-dependent magnetization $M$($H$) of Ti$_4$Ir$_2$O, with the upper critical field identified
from the field-induced features and the field-dependent specific heat $C$/$T$ for Ti$_4$Ir$_2$O, displaying
anomalies near $H_{\rm c2}$ and additional structure at high fields, suggestive of an unconventional high-field superconducting phase.

The effective $g$-factor is a crucial parameter that describes the magnetic moment of a charge carrier. In solid materials, the state of charge carriers is influenced by the electronic band structure, spin-orbit coupling, and the crystal potential. Therefore, the effective $g$-factor in a material is not the same as the free-electron $g$-factor but is modified by the carrier interactions with the crystal environment. As a result, the effective $g$-factor value in a specific material can be very different from $g_{\rm e}$ $\approx$ 2. A reduction of the effective $g$-factor due to spin–orbit coupling and/or the strong-coupling effects with $\Delta_0$ $>$ $\Delta_{BCS}$ or multiple gaps could also result in enhanced $H_{\rm c2}{\rm (0)}$.\cite{altarawneh2012superconducting}

Wu \textit{et al.} provided a possible explanation for the enhanced upper critical field in \ce{Ti4Ir2O} using density functional theory and analytic modeling \cite{wu2025large}. They showed that the nonsymmorphic $Fd\overline{3}m$ symmetry of the $\mathrm{\eta}$-carbide structure enforces strong spin-orbit coupling near the X points of the Brillouin zone, leading to a pronounced suppression of the effective electronic $g$-factor. Such a reduction of the effective $g$-factor naturally weakens Pauli paramagnetic pair breaking and can account for the unusually large upper critical field observed in \ce{Ti4Ir2O}. Experimentally, however, it is challenging to distinguish a genuine violation of the Pauli limit assuming $g = 2$ from an enhancement of the Pauli-limiting field due to a reduced effective $g$-factor, in the absence of direct measurements of the latter \cite{yoshimura2025g}.

Notably, a similarly strong exceedance of the weak-coupling Pauli limit is observed in both \ce{Ti4Co2O} and \ce{Ti4Ir2O}, despite the markedly different atomic weights and therefore the different spin-orbit coupling strengths of Co and Ir.\cite{ma2021group} This observation suggests that the enhancement of the upper critical field may not be governed solely by the atomic spin-orbit coupling of the constituent elements in these materials.

Ruan \textit{et al.} examined whether a reduction of the effective electronic $g$-factor could be responsible for the enhanced upper critical field in \ce{Ti4Ir2O} \cite{ruan2022superconductivity}. By analyzing magnetization data, they estimated the Pauli paramagnetic susceptibility $\chi_{\rm p}$ and extracted a Wilson ratio $R_{\rm W} \approx 3.9$. Such a large Wilson ratio suggests the presence of strong electronic correlations in \ce{Ti4Ir2O}, which may enhance the effective quasiparticle mass and potentially lead to a reduction of the effective $g$-factor. At present there is no direct experimental evidence -- such as from quantum oscillation measurements -- demonstrating that the enhanced upper critical field in \ce{Ti4Ir2O} arises from a reduced effective $g$-factor.

In materials with multiple electronic bands crossing the Fermi level, superconductivity can be more robust against Pauli paramagnetic pair breaking. In this context, Ruan \textit{et al.} reported that a single-gap $s$-wave model fails to reproduce the electronic specific-heat data $C_{\rm e}(T)$ of \ce{Ti4Ir2O} in the superconducting state \cite{ruan2022superconductivity}. Instead, the data are well described by a two-gap $s$-wave model with gap values $\Delta_{0,1} = 1.37$ meV and $\Delta_{0,2} = 0.57$ meV, suggesting possible multiband superconductivity in this compound.

Subsequent muon spin rotation ($\mu$SR) measurements by Das \textit{et al.} reached a different conclusion \cite{das2024ti}. Their analysis of the temperature dependence of the London magnetic penetration depth $\lambda(T)$ is well accounted for by a single-gap $s$-wave model with $\Delta_{0} = 0.92(8)$ meV, providing no clear evidence for multigap superconductivity. In addition, the $\mu$SR measurements revealed that the ratio $T_{\rm c}/\lambda_{\rm eff}^{-2}$ for \ce{Ti4Ir2O} is comparable to values observed in unconventional superconductors \cite{ruan2022superconductivity,das2024ti}.

Taken together, the observation of such large upper critical fields in this material family is particularly striking. $\mathrm{\eta}$-carbide superconductors are cubic, centrosymmetric, and three-dimensional, and therefore lack the low dimensionality or broken inversion symmetry that often accompanies Pauli-limit violation in other systems. The emergence of Pauli-limit-exceeding upper critical fields in these structurally simple, metal-rich compounds is thus unexpected. In addition to their fundamental significance, the exceptionally high upper critical fields observed in $\mathrm{\eta}$-carbide-type superconductors may also be of technological relevance. These compounds are chemically accessible through conventional solid-state synthesis routes, exhibit robust cubic crystal structures, and have upper critical fields as high as commercially used materials.

\section{High pressure measurements on $\mathrm{\eta}$-carbide-type superconductors}

Hydrostatic pressure provides a powerful means to tune crystal structures and electronic properties by continuously modifying interatomic distances and bandwidths.\cite{mao2018solids} In contrast to chemical substitution, pressure offers a clean and reversible approach to optimizing physical properties without introducing disorder, phase separation, or compositional inhomogeneity.

$\mathrm{\eta}$-carbide–type superconductors are hard intermetallic compounds and therefore exhibit high structural robustness under extreme conditions. Recent high-pressure x-ray diffraction measurements have shown that their cubic crystal structures remain stable to pressures of at least 50 GPa \cite{shi2024nonmonotonic}. Under compression, changes in electronic correlations and electron–phonon coupling can significantly influence both the superconducting transition temperature $T_{\rm c}$ and the upper critical field $\mu_{0}H_{\rm c2}(0)$, providing valuable insight into the mechanisms that govern superconductivity in this class of materials. In the following, we review how external pressure can either enhance or suppress $T_{\rm c}$ and $H_{\rm c2}(0)$ in different $\mathrm{\eta}$-carbide–type superconductors.

\begin{figure}[h!]
\centering
\includegraphics[width=\linewidth]{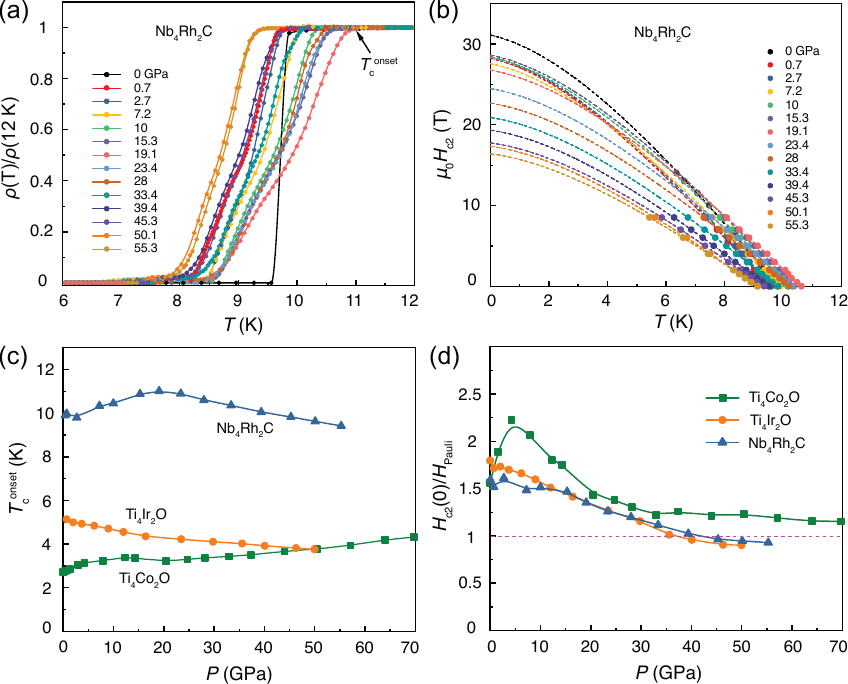}
\caption{Pressure-dependent superconducting properties of $\mathrm{\eta}$-carbide–type compounds. (a) Normalized resistivity $\rho(T)/\rho(12,\mathrm{K})$ of \ce{Nb4Rh2C} under various pressures, illustrating the evolution of the superconducting transition. (b) Temperature dependence of the upper critical field $\mu_{0}H_{\rm c2}(T)$ of \ce{Nb4Rh2C} at selected pressures. (c) Superconducting onset temperature $T_{\rm c}^{\rm onset}$ as a function of pressure for \ce{Nb4Rh2C}, \ce{Ti4Ir2O}, and \ce{Ti4Co2O}. (d) Ratio $\mu_{0}H_{\rm c2}(0)/H_{\rm Pauli}$ versus pressure for the same compounds, showing a crossover from weak-coupling Pauli-limit-exceeding behavior at low pressure to values near or below the weak-coupling Pauli limit at high pressure. Data taken and replotted from \cite{shi2023pressure,shi2024nonmonotonic, Shi2026Ti4Co2O_pressure}.}
\label{fig_pressure}
\end{figure}

Shi \textit{et al.} performed high-pressure studies on \ce{Nb4Rh2C} using x-ray diffraction and electrical transport measurements over a wide pressure range \cite{shi2024nonmonotonic}. They found that the crystal structure remains stable up to at least 58 GPa. Despite this structural robustness, pressure has a pronounced effect on both the superconducting transition temperature $T_{\rm c}$ and the upper critical field $\mu_{0}H_{\rm c2}(0)$, as shown in Figs.~\ref{fig_pressure}\,(a) and (b). In particular, $T_{\rm c}$ exhibits a nonmonotonic pressure dependence, reaching a maximum value of approximately 11 K -- currently the highest $T_{\rm c}$ reported for an $\mathrm{\eta}$-carbide-type superconductor. Additionally, $\mu_{0}H_{\rm c2}(0)$ is progressively reduced with increasing pressure, evolving from well above the Pauli paramagnetic limit at ambient pressure to values below the limit in the range of 40--50 GPa.

In Figs.~\ref{fig_pressure}\,(c) and (d), we summarize the pressure dependence of the superconducting transition temperature and the degree of Pauli-limit violation for \ce{Nb4Rh2C}, \ce{Ti4Ir2O}, and \ce{Ti4Co2O}, highlighting the distinct material-specific responses to external compression.\cite{shi2023pressure,shi2024nonmonotonic, Shi2026Ti4Co2O_pressure}

For \ce{Ti4Ir2O} under pressure, $T_{\rm c}$ decreases gradually from 5.3 K at ambient pressure to 3.8 K at 50 GPa, while x-ray diffraction measurements confirm that the cubic $\mathrm{\eta}$-carbide structure is preserved throughout compression, accompanied only by a reduction of the lattice parameters.\cite{shi2023pressure} The relatively weak pressure dependence of $T_{\rm c}$ was attributed to the strong incompressibility of \ce{Ti4Ir2O}. Interestingly, the zero-temperature upper critical field $\mu_{0}H_{\rm c2}(0)$ undergoes a gradual crossover from values exceeding the weak-coupling BCS Pauli limit to values below the limit at pressures above approximately 35 GPa, as shown in Fig.~\ref{fig_pressure}\,(d).

For \ce{Ti4Co2O}, pressure-induced enhancement of $T_{\rm c}$ has also been reported, which has been attributed to increased electron–phonon coupling and/or modifications of the electronic density of states at the Fermi level. Notably, \ce{Ti4Co2O} has been suggested to undergo a pressure-driven transition between distinct electronic or superconducting states, further underscoring the sensitivity of superconductivity in $\mathrm{\eta}$-carbide–type compounds to external compression. \cite{Shi2026Ti4Co2O_pressure}

\begin{figure}[h]
\centering
\includegraphics[width=\linewidth]{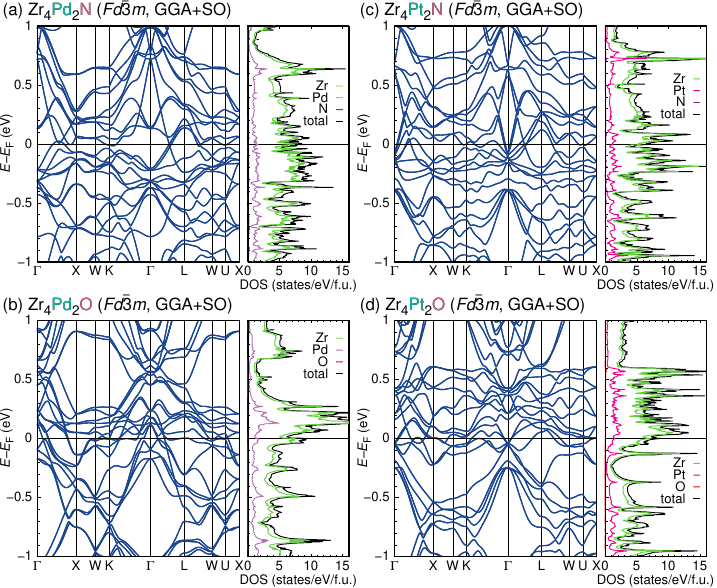}
\caption{Fully relativistic electronic band structures and projected densities of states (DOS) for representative superconducting $\mathrm{\eta}$-carbide–type compounds calculated within GGA including spin–orbit coupling (GGA+SO). Panels (a), (b) show \ce{Zr4Pd2N} and \ce{Zr4Pd2O}, and panels (c), (d) show \ce{Zr4Pt2N} and \ce{Zr4Pt2O}, all in the cubic $Fd\overline{3}m$ structure. Element-resolved DOS highlights the dominant transition-metal contributions near $E_{\rm F}$. }%Data taken and replotted from \cite{}.}
\label{fig:bsdos}
\end{figure}

\begin{figure}[h]
\centering
\includegraphics[width=\linewidth]{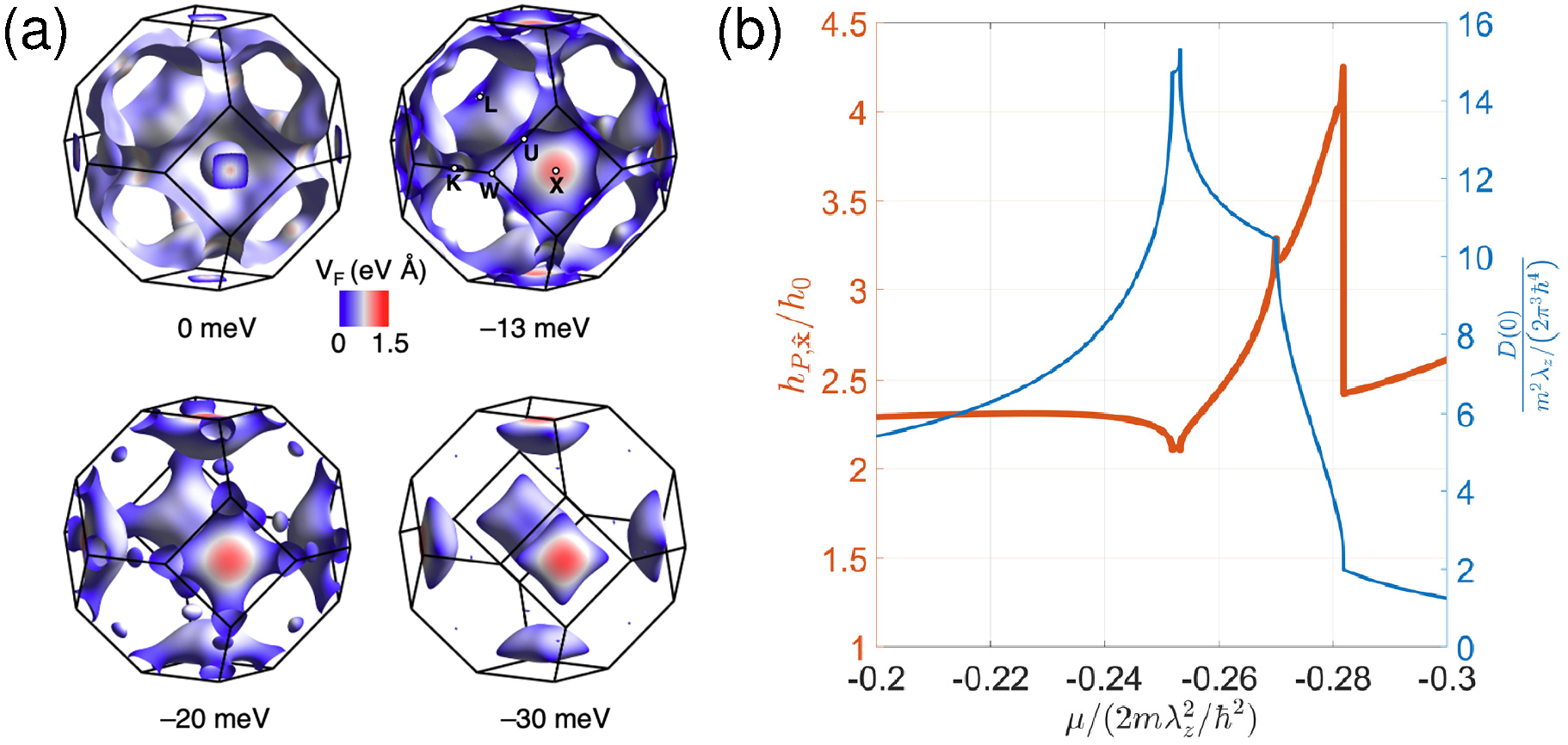}
\caption{(a) Fermi surfaces for four bands of \ce{Ti4Ir2O} at different levels of hole doping. The topology changes as function of chemical potential. (b) Higher order $kp$ theory result for the enhancement of the Pauli field (left scale) and the density of state at the Fermi level as function of the chemical potential (Reprinted with permission from Ref.~\onlinecite{wu2025large}).}
\label{fig:Wu}
\end{figure}

\section{First Principle Calculations and Theory}

The electronic structure of $\mathrm{\eta}$-carbide-type compounds has been repeatedly studied by density functional theory methods. The electronic structure of the prime $\mathrm{\eta}$-carbide examples \ce{W3Fe3C} and \ce{W6Fe6C} was investigated using the Linear muffin-tin orbital method (LMTO)~\cite{ilyasov2008electronic}. Rather unsurprisingly, in $\mathrm{\eta}$-carbide-type compounds with $3d$ and $5d$ transition metals, both $d$ manifolds are partly occupied and pinned to the Fermi level but $3d$ bands are more narrow than $5d$ bands~\cite{Suetin2009}. The  \ce{Fe3W3C}, \ce{Fe6W6C}, \ce{Co3W3C} and \ce{Co6W6C} have been studied in detail using the  full-potential linearized augmented-plane-wave (FLAPW) basis~\cite{Suetin2009}. The electronic structure of \ce{Zr4$M$2O} ($M$ = Fe, Co, Ni) compounds was studied in comparison with X-ray photoelectron spectroscopy (XPS) experiments~\cite{Lavrentyev2013-mb}. Further $\mathrm{\eta}$-carbide structured tungsten carbides and nitrides are studied in Refs.~\onlinecite{Suetin2010,Suetin2016}. DFT was also used extensively to investigate the phase stability of $\mathrm{\eta}$-\ce{Fe2C}~\cite{Fang2012,Ande2012,Oila2014,Lu2018,Yu2024} and of filled \ce{Ti2Ni}-type phases~\cite{lv2015phase,Zhou2022,Ding2025}.

The electronic filling of the \ce{A4B2X} $\mathrm{\eta}$-carbide materials can be modified by shifts in the A, B, and X elements and by off-stoichiometry, in particular in the X content. Shifting from Zr to Nb or from Hf to Ta in the A position adds four electrons to the formula unit. Shifting from Rh to Pd or from Ir to Pt in the B position adds two electrons to the formula unit. And replacing C by N or by O in the X position adds one and two electrons, respectively, to the formula unit. Off-stoichiometry has quite a strong effect in comparison; for X=C, a 16.7\% deficiency removes one electron from the formula unit. This relationship is supported very precisely by cmparison between density of states at the Fermi level $N(E_{\rm F})$ from rigid band theory and $N(E_{\rm F})$ determined from specific heat measurements~\cite{ma2021superconductivity,ma2025discovery}.

The electronic structure of the \ce{A4B2X} type $\mathrm{\eta}$-carbide-type compounds has been studied in some detail for \ce{Ti2Co}, \ce{Ti4Rh2O}, \ce{Ti4Ir2O}~\cite{ma2021group}, for \ce{Nb4Rh2C_{1-$\delta$}}~\cite{ma2021superconductivity} and for \ce{Ta4Rh2C_{1-$\delta$}}~\cite{ma2025discovery}. Fig.~\ref{fig:bsdos} shows four examples of electronic structures of superconducting $\mathrm{\eta}$-carbide-type compounds (see also Ref.~\onlinecite{Gupta2009} for the DOS of \ce{Zr4Pt2O}). They are calculated with the full potential local orbital (FPLO) basis~\cite{Koepernik1999} in combination with a generalized gradient approximation (GGA) exchange correlation potential~\cite{Perdew1996}. We employ fully relativistic GGA that includes the effects of spin-orbit coupling (SOC); this is relevant because of the $4d$ and $5d$ transition metals. The effects of SOC on the bands are significantly stronger for the $5d$ compounds~\cite{ma2025discovery}. As the primitive cell of the $Fd\bar{3}m$ \ce{A4B2X} materials contains $Z=4$ formula units, {\it i.e.} 28 atoms, 24 of which are transition metals, the electronic structure appears rather complex, with 240 $d$ bands near the Fermi level. This also leads to complex Fermi surfaces~\cite{ma2021group}. The four compounds \ce{Zr4Pd2N}, \ce{Zr4Pd2O}, \ce{Zr4Pt2N} and \ce{Zr4Pt2O} represent the isoelectronic substitution of Pd by Pt (Figs.~\ref{fig:bsdos}\,(a) and (c), Figs.~\ref{fig:bsdos}\,(b) and (d), respectively). For the small energy range shown here, the increase of SOC and the differences in interatomic distances lead to a substantial change of electronic structure so that the isoelectronic nature is not obvious. The addition of a single electron to the formula unit by the replacement of N by O is seen in  Figs.~\ref{fig:bsdos}\,(a) and (b), Figs.~\ref{fig:bsdos}\,(c) and (d), respectively. Addition of one electron in rigid band approximation would shift the Fermi level up by 0.14\,eV for \ce{Zr4Pd2N} or by 0.13\,eV for \ce{Zr4Pt2N} but the real change in electronic structure is far more substantial. Overall, the four examples show that the weight of Zr dominates at the Fermi level, even somewhat exceeding the weight of Zr in the stoichiometry. Pd and Pt also contribute to the DOS at the Fermi level, but N and O are negligible.

As mentioned above, the large upper critical field observed in some $\mathrm{\eta}$-carbide superconductors like \ce{Zr4Pd2O}~\cite{watanabe2023observation}, \ce{Nb4Rh2C_{1-$\delta$}}~\cite{ma2021superconductivity} and \ce{Ti4Ir2O}~\cite{ma2021group} is studied theoretically in Ref.~\onlinecite{wu2025large}. Focusing on the electronic structure of \ce{Ti4Ir2O}, the authors observe that the Fermi surface is dominated by a pocket at the X point, especially for certain values of hole doping (see Fig.~\ref{fig:Wu}\,(a)). They apply $kp$ theory to model this pocket. While the simple elliptical pocket is not relevant for the material \ce{Ti4Ir2O}, the authors find a sharp enhancement of the Pauli critical field for particular values of the chemical potential. The cause of the enhancement is a local suppression of the effective $g$-factor as the logarithm of this quantity is averaged over the Brillouin zone to obtain the Pauli limiting field. For \ce{Ti4Ir2O}, the authors apply a higher order $kp$ theory and find the Pauli limiting field shown in Fig.~\ref{fig:Wu}\,(b). For the considered (hole doping) chemical potentials, the Pauli field is always above 2 (also see discussion above).  

\section{Conclusion and Outlook}

In this review, we have surveyed recent progress on superconductivity in $\mathrm{\eta}$-carbide–type compounds, a family of metal-dense materials that combine high crystallographic symmetry with intricate multiband electronic structures. $\mathrm{\eta}$-Carbide-type compounds have recently emerged as a distinct class of superconductors characterized by wide compositional flexibility and systematic chemical tunability. Bulk superconductivity in these materials has now been firmly established through transport, magnetic, and thermodynamic measurements. Investigations on phase-pure compounds have resolved ambiguities in the literature and defined a growing set of well-characterized superconductors, with transition temperatures approaching 10 K and, in several cases, exceptionally large upper critical fields.

A central finding highlighted throughout this review is the prevalence of exceptionally high upper critical fields in several $\mathrm{\eta}$-carbide superconductors, even exceeding the weak-coupling BCS Pauli paramagnetic limit. This behavior is surprising given the centrosymmetric, three-dimensional, and structurally isotropic nature of these compounds. Experimental studies reveal that Pauli-limit violation is a robust and reproducible feature in multiple members of the family, while high-pressure measurements demonstrate that both the superconducting transition temperature and the degree of Pauli-limit exceedance can be continuously tuned without structural phase transitions. The systematic suppression of Pauli-limit violation under pressure further indicates that the high-field behavior is intimately linked to the electronic structure rather than to extrinsic disorder or structural instabilities.

Density-functional studies reveal complex, multiband electronic structures with strong spin–orbit coupling effects arising from the nonsymmorphic $Fd\overline{3}m$ symmetry and the presence of heavy transition-metal elements. Recent theoretical work has shown that symmetry-protected spin-orbit-coupled states near the Fermi level can strongly renormalize the effective electronic $g$-factor, offering a natural explanation for the enhanced Pauli-limiting field observed in compounds such as \ce{Ti4Ir2O}. At the same time, experimental indications of multiband superconductivity, enhanced Wilson ratios, and anomalous thermodynamic responses suggest that strong coupling and correlation effects may also play an important role.

Looking forward, several directions appear particularly promising. First, direct experimental probes of the effective $g$-factor, such as quantum-oscillation or spin-resolved spectroscopic measurements, are essential to establish whether Pauli-limit exceedance in this family is governed primarily by spin–orbit–driven renormalization or by alternative mechanisms. Second, further high-field thermodynamic and transport studies are needed to assess the possible emergence of exotic high-field superconducting states, including the proposed FFLO finite-momentum pairing. Third, the remarkable compositional variability of the $\mathrm{\eta}$-carbide framework offers a platform for materials design. This may include the controlled tuning of electron count, transition-metal species, and light-element occupancy which may provide access to band filling and spin-orbit coupling strength.

\section*{Acknowledgments}

H.~O.~J. acknowledges support through JSPS KAKENHI Grants No.~24H01668 and No.~25K0846007. This work was supported by the Swiss National Science Foundation under Grant No. PCEFP2\_194183. We thank Hiroyuki Nakamura for insightful discussions. We thank Yuto Watanabe and Yoshikazu Mizuguchi for providing the original measurement data of \ce{Zr4Pd2O} for figure \ref{fig:supercon}. We thank R. Lortz for providing the original vector graphics of the figure \ref{fig:highfield}(c) and (d).

\bibliography{References}% Produces the bibliography via BibTeX.

@article{ku1984new,
  title={New superconducting ternary transition metal compounds with the {E}9$_3$-type structure},
  author={Ku, HC and Johnston, DC},
  xjournal={Chinese Journal of Physics},
  journal={Chin. J. Phys.},
  volume={22},
  number={1},
  pages={59--64},
  year={1984},
  publisher={台灣物理學會}
}

@book{poole1999handbook,
  title={Handbook of superconductivity},
  author={Poole, Charles K and Farach, Horacio A and Creswick, Richard J},
  year={1999},
  publisher={Elsevier}
}

@article{ku1985effect,
  title={Effect of composition on the superconductivity of the {E}9$_3$ phase in the ternary \ce{Nb-Rh-C} system},
  author={Ku, HC},
  journal={Physica B+C},
  volume={135},
  number={1-3},
  pages={417--419},
  year={1985},
  publisher={Elsevier}
}

@article{ma2021superconductivity,
  title={Superconductivity with high upper critical field in the cubic centrosymmetric $\eta$-carbide \ce{Nb4Rh2C_{1-\delta}}},
  author={Ma, KeYuan and Gornicka, Karolina and Lef{\`e}vre, Robin and Yang, Yikai and R{\o}nnow, Henrik M and Jeschke, Harald O and Klimczuk, Tomasz and von Rohr, Fabian O},
  journal={ACS Materials Au},
  volume={1},
  number={1},
  pages={55--61},
  year={2021},
  publisher={ACS Publications}
}

@article{ma2025discovery,
  title={Discovery of the type-{II} superconductor \ce{Ta4Rh2C_{1-\delta}} with a high upper critical field},
  author={Ma, KeYuan and L{\'o}pez-Paz, Sara and Gornicka, Karolina and Jeschke, Harald O and Klimczuk, Tomasz and von Rohr, Fabian O},
  xjournal={Physical Review Research},
  journal={Phys. Rev. Res.},
  volume={7},
  number={2},
  pages={023147},
  year={2025},
  publisher={APS}
}

@article{ma2021group,
  title={Group-9 transition-metal suboxides adopting the filled-\ce{Ti2Ni} structure: A class of superconductors exhibiting exceptionally high upper critical fields},
  author={Ma, KeYuan and Lef{\`e}vre, Robin and Gornicka, Karolina and Jeschke, Harald O and Zhang, Xiaofu and Guguchia, Zurab and Klimczuk, Tomasz and von Rohr, Fabian O},
  xjournal={Chemistry of Materials},
  journal={Chem. Mater.},
  volume={33},
  number={22},
  pages={8722--8732},
  year={2021},
  publisher={ACS Publications}
}

@article{das2024ti,
  title={\ce{Ti4Ir2O}: A time reversal invariant fully gapped unconventional superconductor},
  author={Das, Debarchan and Ma, KeYuan and Jaroszynski, Jan and Sazgari, Vahid and Klimczuk, Tomasz and von Rohr, Fabian O and Guguchia, Zurab},
  xjournal={Physical Review B},
  journal={Phys. Rev. B},
  volume={110},
  number={17},
  pages={174507},
  year={2024},
  publisher={APS}
}

@article{ruan2022superconductivity,
  title={Superconductivity with a violation of {P}auli limit and evidences for multigap in $\eta$-carbide type \ce{Ti4Ir2O}},
  author={Ruan, Bin-Bin and Zhou, Meng-Hu and Yang, Qing-Song and Gu, Ya-Dong and Ma, Ming-Wei and Chen, Gen-Fu and Ren, Zhi-An},
  xjournal={Chinese Physics Letters},
  journal={Chin. Phys. Lett.},
  volume={39},
  number={2},
  pages={027401},
  year={2022},
  publisher={IOP Publishing}
}

@article{ma2019superconductivity,
  title={Superconductivity in the $\eta$-carbide-type oxides \ce{Zr4Rh2O_x}},
  author={Ma, KeYuan and Lago, Jorge and von Rohr, Fabian O},
  xjournal={Journal of Alloys and Compounds},
  journal   = "J. Alloys Compd.",
  volume={796},
  pages={287--292},
  year={2019},
  publisher={Elsevier}
}

@ARTICLE{Waki2011,
  title     = "Interplay between quantum criticality and geometric frustration
               in \ce{Fe3Mo3N} with stella quadrangula lattice",
  author    = "Waki, T and Terazawa, S and Yamazaki, T and Tabata, Y and Sato, K
               and Kondo, A and Kindo, K and Yokoyama, M and Takahashi, Y and
               Nakamura, H",
  journal   = "Europhys. Lett.",
  publisher = "IOP Publishing",
  volume    =  94,
  number    =  3,
  pages     =  37004,
  month     =  may,
  year      =  2011,
}

@ARTICLE{nevitt1960,
  title   = "A Further Study of \ce{Ti2Ni}-Type Phases Containing Titanium,
             Zirconium or Hafnium",
  author  = "{M. V. Nevitt, J. W. Downey, R. A. Morris}",
  xjournal = "Trans. Soc. Pet. Eng. Am. Inst. Min. Metall. Pet. Eng. Inc",
  journal   = "Trans. Metall. Soc. AIME",
  volume    =  218,
  pages="1019",
  year    =  1960
}

@ARTICLE{Mackay1994,
  title     = "New oxides of the filled-\ce{Ti2Ni} type structure",
  author    = "Mackay, Richard and Miller, Gordon J and Franzen, Hugo F",
  journal   = "J. Alloys Compd.",
  publisher = "Elsevier BV",
  volume    =  204,
  number    = "1-2",
  pages     = "109--118",
  month     =  feb,
  year      =  1994,
}

@ARTICLE{Cantrell2002,
  title     = "{X}-ray diffraction, neutron scattering and {NMR} studies of hydrides formed by \ce{Ti4Pd2O} and \ce{Zr4Pd2O}",
  author    = "Cantrell, J S and Bowman, Jr, R C and Maeland, A J",
  journal   = "J. Alloys Compd.",
  publisher = "Elsevier BV",
  volume    = "330-332",
  pages     = "191--196",
  month     =  jan,
  year      =  2002,
}

@article{watanabe2023observation,
  title={Observation of superconductivity and enhanced upper critical field of $\eta$-carbide-type oxide \ce{Zr4Pd2O}},
  author={Watanabe, Yuto and Miura, Akira and Moriyoshi, Chikako and Yamashita, Aichi and Mizuguchi, Yoshikazu},
  journal={Sci. Rep.},
  volume={13},
  number={1},
  pages={22458},
  year={2023},
  publisher={Nature Publishing Group UK London}
}

@ARTICLE{Weil1997,
  title     = "Synthesis of a new ternary nitride, \ce{Fe4W2N}, with a
               unique $\eta$-carbide structure",
  author    = "Weil, K S and Kumta, P N",
  journal   = "J. Solid State Chem.",
  publisher = "Elsevier BV",
  volume    =  134,
  number    =  2,
  pages     = "302--311",
  month     =  dec,
  year      =  1997,
}

@ARTICLE{Kotyk1970,
  title     = {Study of filled \ce{Ti2Ni}-type phases with hafnium, tantalum, and tungsten},
  author    = "Kotyk, M and Stadelmaier, H H",
  journal   = "Metall. Trans.",
  publisher = "Springer Science and Business Media LLC",
  volume    =  1,
  number    =  4,
  pages     = "899--903",
  month     =  apr,
  year      =  1970,
}

@ARTICLE{Kuo1953,
  title     = "The formation of $\eta$ carbides",
  author    = "Kuo, Kehsin",
  journal   = "Acta Metall.",
  publisher = "Elsevier BV",
  volume    =  1,
  number    =  3,
  pages     = "301--304",
  month     =  may,
  year      =  1953,
}

@ARTICLE{Leonard1991,
  title     = "Structure determinations of two new ternary oxides: {Ti3PdO} and
               {Ti4Pd2O}",
  author    = "Leonard, Susan R and Snyder, Barry S and Brewer, Leo and Stacy,
               Angelica M",
  journal   = "J. Solid State Chem.",
  publisher = "Elsevier BV",
  volume    =  92,
  number    =  1,
  pages     = "39--50",
  month     =  may,
  year      =  1991,
}

@article{bojarski1967neutron,
  title={NEUTRON DIFFRACTION STUDY OF THE CRYSTAL STRUCTURES OF ETA-PHASE IRON--TUNGSTEN CARBIDES.},
  author={Bojarski, Z and Leciejewicz, J},
  journal={Arch. Hutnictwa},
  volume ={12},
  pages = {255},
  year={1967}
}

@PHDTHESIS{Ma2022,
  title    = "Superconductors with an $\eta$-carbide type structure: a class of
              superconductors exhibiting exceptionally high upper critical
              fields",
  author   = "Ma, Keyuan",
  address  = "Zurich",
  year     =  2022,
  school   = "University of Zurich"
}

@misc{Waki2012Zr3V3O,
  author       = {Waki, T. and Inoue, T. and Tabata, Y. and Nakamura, H.},
  title        = {Superconductivity in $\eta$-carbide-type oxide \ce{Zr3V3O}},
  howpublished = {Talk presented at the Japan Society of Powder and Powder Metallurgy Spring Meeting},
  year         = {2012},
  note         = {{M}ay 24, 2012}
}

@article{hirotsu1972crystal,
  title={Crystal structure and morphology of the carbide precipitated from martensitic high carbon steel during the first stage of tempering},
  author={Hirotsu, Y and Nagakura, S},
  journal={Acta Metallurgica},
  volume={20},
  number={4},
  pages={645--655},
  year={1972},
  publisher={Elsevier}
}

@article{jack1973invited,
  title={Invited review: carbides and nitrides in steel},
  author={Jack, DH and Jack, KH},
  journal={Materials Science and Engineering},
  volume={11},
  number={1},
  pages={1--27},
  year={1973},
  publisher={Elsevier}
}

@BOOK{Toth1971,
  title     = "Transition Metal Carbides and Nitrides",
  author    = "Toth, Louis E",
  editor    = "Margrave, John L",
  publisher = "Academic Press Inc. (London)",
  address   = "London, England",
  year      =  1971
}

@ARTICLE{Mueller1963,
  title     = "The Crystal Structures of \ce{Ti2Cu}, \ce{Ti2Ni}, \ce{Ti4Ni2O}, and
               \ce{Ti4Cu2O}",
  author    = "Mueller, M and Knott, H W",
  xjournal   = "Transactions of American Institute of Metallurgical Engineers",
  journal   = "Trans. Metall. Soc. AIME",
  pages="674",
  publisher = "Argonne National Lab., Ill.; Argonne National Lab.(ANL), Argonne,
               IL (United …",
  volume    =  227,
  number    = "ANL-FGF-396",
  month     =  jun,
  year      =  1963,
}

@ARTICLE{Pollock1970,
  title     = "The eta carbides in the {Fe-W-C} and {Co-W-C} systems",
  author    = "Pollock, C B and Stadelmaier, H H",
  journal   = "Metall. Trans.",
  publisher = "Springer Science and Business Media LLC",
  volume    =  1,
  number    =  4,
  pages     = "767--770",
  month     =  apr,
  year      =  1970,
}

@ARTICLE{Zavaliy1999,
  title     = "Effect of oxygen content on hydrogen storage capacity of \ce{Zr}-based
               $\eta$-phases",
  author    = "Zavaliy, I Yu",
  journal   = "J. Alloys Compd.",
  publisher = "Elsevier BV",
  volume    =  291,
  number    = "1-2",
  pages     = "102--109",
  month     =  sep,
  year      =  1999,
}

@ARTICLE{Lavrentyev2013-mb,
  title   = "Electronic properties of \ce{ZrMO} (\textit{M} $=$ \ce{Fe, Co, Ni}) intermetallic
             compounds: first-principles {APW}+{LO} calculations and {X}-ray
             photoelectron spectroscopy data",
  author  = "Lavrentyev, A A and Gabrelian, B V and Shkumat, P N and Kopylova, E
             I and Nikiforov, I Y and Zavaliy, I Y and Sinelnichenko, A K and
             Khyzhun, O Y",
  journal = "Chem. Met. Alloys",
  volume  =  6,
  pages   = "150--157",
  year    =  2013
}

@ARTICLE{Holleck1967,
  title   = "{Ternäre {K}omplex-carbide, -nitride und -oxide mit teilweise
             aufgefüllter \ce{Ti2Ni}-{S}truktur}",
  author  = "Holleck, H and Thümmler, F",
  xjournal = "Monatshefte für Chemie und verwandte Teile anderer Wissenschaften",
journal={Monatsh. Chem.},
  volume  =  98,
  pages   = "133--134",
  year    =  1967
}

@ARTICLE{Holleck1967nuc,
  title     = "{Untersuchungen über die {B}ildung von nichtmetallstabilisierten zirkonreichen {\"U}bergangsmetallphasen}",
  author    = "{Holleck, H and Th{\"u}mmler, F}",
  journal   = "J. Nucl. Mater.",
  publisher = "Elsevier BV",
  volume    =  23,
  number    =  1,
  pages     = "88--94",
  month     =  jul,
  year      =  1967,
}

@misc{hu2023thermodynamic,
  title={Thermodynamic signatures of a potential {F}ulde-{F}errell-{L}arkin {O}vchinnikov state in the isotropic superconductor \ce{Ti4Ir2O}},
  author={Hu, Jiaqi and Hei Ng, Yat and Atanov, Omargeldi and Ruan, Bin-Bin and Ren, Zhi-An and Lortz, Rolf},
      year={2024},
      eprint={2312.01914},
      archivePrefix={arXiv},
      primaryClass={cond-mat.supr-con},
      url={https://arxiv.org/abs/2312.01914} 
}

@ARTICLE{Nyman1978,
  title     = "The pyrochlore structure and its relatives",
  author    = "Nyman, Harry and Andersson, Sten and Hyde, B G and O'Keeffe, M",
  journal   = "J. Solid State Chem.",
  publisher = "Elsevier BV",
  volume    =  26,
  number    =  2,
  pages     = "123--131",
  month     =  oct,
  year      =  1978,
}

@ARTICLE{Vandenberg1976,
  title     = "Superconductivity of a new metastable phase of scandium-chromium",
  author    = "Vandenberg, J M and Matthias, B T and Corenzwit, E and Barz, H",
  journal   = "J. Solid State Chem.",
  publisher = "Elsevier BV",
  volume    =  18,
  number    =  4,
  pages     = "395--396",
  month     =  aug,
  year      =  1976,
}

@ARTICLE{Souissi2018,
  title     = "Effect of mixed partial occupation of metal sites on the phase
               stability of $\gamma$-\ce{Cr_{23-x}Fe_{x}C6} ($x = 0$-$3$) carbides",
  author    = "Souissi, Maaouia and Sluiter, Marcel H F and Matsunaga, Tetsuya
               and Tabuchi, Masaaki and Mills, Michael J and Sahara, Ryoji",
  journal   = "Sci. Rep.",
  publisher = "Nature Publishing Group",
  volume    =  8,
  number    =  1,
  pages     =  7279,
  month     =  may,
  year      =  2018,
}

@ARTICLE{Westgren1926,
  title     = "{R\"ontgenanalyse der {S}ysteme {W}olfram‐{K}ohlenstoff und
               {M}olybd\"an‐{K}ohlenstoff}",
  author    = "Westgren, Arne and Phragmén, Gösta",
  journal   = "Z. Allg. Anorg. Chem.",
  publisher = "Wiley",
  volume    =  156,
  number    =  1,
  pages     = "27--36",
  month     =  sep,
  year      =  1926,
}

@ARTICLE{Karlsson1951,
  title     = "Metallic oxides with the structure of high-speed steel carbide",
  author    = "Karlsson, Nils",
  journal   = "Nature",
  publisher = "Springer Science and Business Media LLC",
  volume    =  168,
  number    =  4274,
  pages     = "558--558",

  month     =  sep,
  year      =  1951,
}

@ARTICLE{Nohara2024,
  title     = "Metal-rich compounds: A new platform for superconductivity
               research",
  author    = "Nohara, Minoru",
  journal   = "JPSJ News Comments",
  publisher = "Physical Society of Japan",
  volume    =  21,
  number    =  01,
  month     =  jan,
  year      =  2024,
}

@article{parthe1965neutron,
  title={A neutron diffraction study of the {N}owotny phase \ce{Mo45Si3C41}},
  author={Parthe, E and Jeitschko, WOLFGANG and Sadagopan, VARADACHARI},
  xjournal={Acta Crystallographica},
  journal={Acta Cryst.},
  volume={19},
  number={6},
  pages={1031--1037},
  year={1965},
  publisher={International Union of Crystallography}
}

@ARTICLE{Westgren1933,
  title     = "Complex chromium and iron carbides",
  author    = "Westgren, A",
  journal   = "Nature",
  publisher = "Springer Science and Business Media LLC",
  volume    =  132,
  number    =  3334,
  pages     = "480--480",
  month     =  sep,
  year      =  1933,
}

@ARTICLE{Waki2010,
  title     = "Non-{F}ermi-liquid behavior on an iron-based itinerant electron
               magnet \ce{Fe3Mo3N}",
  author    = "Waki, Takeshi and Terazawa, Shinsuke and Tabata, Yoshikazu and
               Oba, Fumiyasu and Michioka, Chishiro and Yoshimura, Kazuyoshi and
               Ikeda, Shugo and Kobayashi, Hisao and Ohoyama, Kenji and
               Nakamura, Hiroyuki",
  journal   = "J. Phys. Soc. Jpn.",
  publisher = "Physical Society of Japan",
  volume    =  79,
  number    =  4,
  pages     =  043701,
  month     =  mar,
  year      =  2010,
}

@ARTICLE{Liu2024,
  title     = "Crystal structure of \ce{Ti4Ni2C}",
  author    = "Liu, Huizi and Liang, Xinyu and Liu, Yibo and Fan, Changzeng and
               Wen, Bin and Zhang, Lifeng",
  journal   = "IUCrdata",
  publisher = "International Union of Crystallography (IUCr)",
  volume    =  9,
  number    = "Pt 1",
  pages     = "x240043",
  month     =  jan,
  year      =  2024,
}

@article{taylor1952new,
  title={A new complex eta-carbide},
  author={Taylor, A and Sachs, K},
  journal={Nature},
  volume={169},
  number={4297},
  pages={411--411},
  year={1952},
  publisher={Nature Publishing Group UK London}
}

@Article{Jeitschko1964,
author={Jeitschko, W.
and Holleck, H.
and Nowotny, H.
and Benesovsky, F.},
title={Phasen mit aufgef{\"u}lltem \ce{Ti2Ni}-Typ},
xjournal={Monatshefte f{\"u}r Chemie und verwandte Teile anderer Wissenschaften},
journal={Monatsh. Chem.},
year={1964},
month={May},
day={01},
volume={95},
number={3},
pages={1004-1006},
issn={1434-4475},
doi={10.1007/BF00908814},
url={https://doi.org/10.1007/BF00908814}
}

@article{Gupta2009,
title = {Structural and compositional investigations of \ce{Zr4Pt2O}: A filled-cubic \ce{Ti2Ni}-type phase},
xjournal = {Journal of Solid State Chemistry},
journal = {J. Solid State Chem.},
volume = {182},
number = {7},
pages = {1708-1712},
year = {2009},
issn = {0022-4596},
doi = {https://doi.org/10.1016/j.jssc.2009.04.011},
url = {https://www.sciencedirect.com/science/article/pii/S0022459609001674},
author = {Shalabh Gupta and Daniel J. Sordelet and John D. Corbett}
}

@article{mao2018solids,
  title={Solids, liquids, and gases under high pressure},
  author={Mao, Ho-Kwang and Chen, Xiao-Jia and Ding, Yang and Li, Bing and Wang, Lin},
  xjournal={Reviews of Modern Physics},
  journal={Rev. Mod. Phys.},
  volume={90},
  number={1},
  pages={015007},
  year={2018},
  publisher={APS}
}

@article{kinnunen2018fulde,
  title={The {F}ulde--{F}errell--{L}arkin--{O}vchinnikov state for ultracold fermions in lattice and harmonic potentials: a review},
  author={Kinnunen, Jami J and Baarsma, Jildou E and Martikainen, Jani-Petri and T{\"o}rm{\"a}, P{\"a}ivi},
  xjournal={Reports on Progress in Physics},
  journal={Rep. Prog. Phys.},
  volume={81},
  number={4},
  pages={046401},
  year={2018},
  publisher={IOP Publishing}
}

@book{tinkham2004introduction,
  title={Introduction to superconductivity},
  author={Tinkham, Michael},
  year={2004},
  publisher={Courier Corporation}
}

@article{kirshenbaum2013pressure,
  title={Pressure-Induced Unconventional Superconducting Phase in the Topological Insulator \ce{Bi2Se3}},
  author={Kirshenbaum, Kevin and Syers, PS and Hope, AP and Butch, NP and Jeffries, JR and Weir, ST and Hamlin, JJ and Maple, MB and Vohra, YK and Paglione, J},
  xjournal={Physical Review Letters},
  journal={Phys. Rev. Lett.},
  volume={111},
  number={8},
  pages={087001},
  year={2013},
  publisher={American Physical Society (APS)}
}

@article{maki1966effect,
  title={Effect of Pauli paramagnetism on magnetic properties of high-field superconductors},
  author={Maki, Kazumi},
  xjournal={Physical Review},
  journal={Phys. Rev.},
  volume={148},
  number={1},
  pages={362},
  year={1966},
  publisher={APS}
}

@article{clogston1962upper,
  title={Upper limit for the critical field in hard superconductors},
  author={Clogston, Albert M},
  xjournal={Physical Review Letters},
  journal={Phys. Rev. Lett.},
  volume={9},
  number={6},
  pages={266},
  year={1962},
  publisher={APS}
}

@article{Lu2018,
title = {Formation of eta carbide in ferrous martensite by room temperature aging},
xjournal = {Acta Materialia},
journal = {Acta Mater.},
volume = {158},
pages = {297-312},
year = {2018},
issn = {1359-6454},
doi = {https://doi.org/10.1016/j.actamat.2018.07.071},
url = {https://www.sciencedirect.com/science/article/pii/S1359645418306128},
author = {W. Lu and M. Herbig and C.H. Liebscher and L. Morsdorf and R.K.W. Marceau and G. Dehm and D. Raabe}
}

@Article{Oila2014,
author={Oila, Adrian
and Lung, Chi
and Bull, Steve},
title={Elastic properties of eta carbide ($\eta$-\ce{Fe2C}) from ab initio calculations: application to cryogenically treated gear steel},
xjournal={Journal of Materials Science},
journal={J. Mater. Sci.},
year={2014},
month={Mar},
day={01},
volume={49},
number={5},
pages={2383-2390},
issn={1573-4803},
doi={10.1007/s10853-013-7942-0},
url={https://doi.org/10.1007/s10853-013-7942-0}
}

@article{Ding2025,
title = {High-temperature structural stability and mechanical properties of $\eta$-carbides \ce{M6W6C}, \ce{M3W3C} and \ce{M2W2C} from first-principles calculations},
xjournal = {Ceramics International},
journal = {Ceram. Intl.},
volume = {51},
number = {16, Part A},
pages = {21742-21751},
year = {2025},
issn = {0272-8842},
doi = {10.1016/j.ceramint.2025.02.335},
xurl = {https://www.sciencedirect.com/science/article/pii/S0272884225010181},
author = {Juan Ding and Yunzhu Ma and Wensheng Liu and Chaoping Liang}
}

@article{Suetin2009,
title = {Structural, electronic and magnetic properties of $\eta$ carbides (\ce{Fe3W3C}, \ce{Fe6W6C}, \ce{Co3W3C} and \ce{Co6W6C}) from first principles calculations},
xjournal = {Physica B: Condensed Matter},
journal = {Physica B},
volume = {404},
number = {20},
pages = {3544-3549},
year = {2009},
issn = {0921-4526},
doi = {10.1016/j.physb.2009.05.051},
xurl = {https://www.sciencedirect.com/science/article/pii/S0921452609003421},
author = {D.V. Suetin and I.R. Shein and A.L. Ivanovskii}
}

@article{Suetin2010,
doi = {10.1070/RC2010v079n07ABEH004141},
url = {https://dx.doi.org/10.1070/RC2010v079n07ABEH004141},
year = {2010},
month = {sep},
publisher = {},
volume = {79},
number = {7},
pages = {611},
author = {D V Suetin and I R Shein and Alexander L Ivanovskii},
title = {Tungsten carbides and nitrides and ternary systems based on them: 
the electronic structure, chemical bonding and properties},
xjournal = {Russian Chemical Reviews},
journal = {Russ. Chem. Rev.}
}

@article{Zhou2022,
title = {Lattice stability, mechanical and thermal properties of a new class of multicomponent \ce{(Fe, Mo, W)6C} $\eta$ carbides with different atomic site configurations},
xjournal = {Ceramics International},
journal = {Ceram. Intl.},
volume = {48},
number = {4},
pages = {5107-5118},
year = {2022},
issn = {0272-8842},
doi = {10.1016/j.ceramint.2021.11.049},
url = {https://www.sciencedirect.com/science/article/pii/S0272884221034568},
author = {Yunxuan Zhou and Yang Lin and Fei Zhang and Yehua Jiang and Shizhong Wei and Liujie Xu and Xiaoyu Chong and Zulai Li and Jing Feng}
}

@article{Suetin2016,
title = {Structural, electronic and magnetic properties of $\eta$-carbides \ce{$M$3W3C} (${M}$= {Ti}, {V}, {Cr}, {Mn}, {Fe}, {Co}, {Ni})},
xjournal = {Journal of Alloys and Compounds},
  journal   = "J. Alloys Compd.",
volume = {681},
pages = {508-515},
year = {2016},
issn = {0925-8388},
doi = {https://doi.org/10.1016/j.jallcom.2016.04.279},
url = {https://www.sciencedirect.com/science/article/pii/S092583881631266X},
author = {D.V. Suetin and N.I. Medvedeva}
}

@Article{Ande2012,
author={Ande, Chaitanya Krishna
and Sluiter, Marcel H. F.},
title={First-Principles Calculations on Stabilization of Iron Carbides (\ce{Fe3C}, \ce{Fe5C2}, and $\eta$-\ce{Fe2C}) in Steels by Common Alloying Elements},
xjournal={Metallurgical and Materials Transactions A},
journal={Metall. Mater. Trans. A},
year={2012},
month={Nov},
day={01},
volume={43},
number={11},
pages={4436-4444},
issn={1543-1940},
doi={10.1007/s11661-012-1229-y},
url={https://doi.org/10.1007/s11661-012-1229-y}
}

@article{ilyasov2008electronic,
  title={Electronic structure and chemical bond in carbides crystallizing in the \ce{Fe-W-C} system},
  author={Ilyasov, AV and Ryzhkin, AA and Ilyasov, VV},
  xjournal={Journal of Structural Chemistry},
  journal={J. Struct. Chem.},
  volume={49},
  number={5},
  pages={795--802},
  year={2008},
  publisher={Springer}
}

@article{lv2015phase,
  title={Phase stability, electronic and elastic properties of \ce{Fe_{6-x}W_{x}C} ($x= 0$-$6$) from density functional theory},
  author={Lv, ZQ and Zhou, ZA and Sun, SH and Fu, WT},
  xjournal={Materials Chemistry and Physics},
  journal={Mater. Chem. Phys.},
  volume={164},
  pages={115--121},
  year={2015},
  publisher={Elsevier}
}

@article{ganin2010polymorphism,
  title={Polymorphism control of superconductivity and magnetism in \ce{Cs3C60} close to the {M}ott transition},
  author={Ganin, Alexey Y and Takabayashi, Yasuhiro and Jegli{\v{c}}, Peter and Ar{\v{c}}on, Denis and Poto{\v{c}}nik, Anton and Baker, Peter J and Ohishi, Yasuo and McDonald, Martin T and Tzirakis, Manolis D and McLennan, Alec and others},
  journal={Nature},
  volume={466},
  number={7303},
  pages={221--225},
  year={2010},
  publisher={Nature Publishing Group UK London}
}

@article{ganin2008bulk,
  title={Bulk superconductivity at 38 {K} in a molecular system},
  author={Ganin, Alexey Y and Takabayashi, Yasuhiro and Khimyak, Yaroslav Z and Margadonna, Serena and Tamai, Anna and Rosseinsky, Matthew J and Prassides, Kosmas},
  xjournal={Nature materials},
  journal={Nat. Mater.},
  volume={7},
  number={5},
  pages={367--371},
  year={2008},
  publisher={Nature Publishing Group UK London}
}

@article{weller2005superconductivity,
  title={Superconductivity in the intercalated graphite compounds \ce{C6Yb} and \ce{C6Ca}},
  author={Weller, Thomas E and Ellerby, Mark and Saxena, Siddharth S and Smith, Robert P and Skipper, Neal T},
  xjournal={Nature Physics},
  journal={Nat. Phys.},
  volume={1},
  number={1},
  pages={39--41},
  year={2005},
  publisher={Nature Publishing Group UK London}
}

@article{mazin2005intercalant,
  title={Intercalant-driven superconductivity in \ce{YbC6} and \ce{CaC6}},
  author={Mazin, II},
  xjournal={Physical Review Letters},
  journal={Phys. Rev. Lett.},
  volume={95},
  number={22},
  pages={227001--227001},
  year={2005}
}

@incollection{kobayashi2019superconductivity,
  title={Superconductivity of carbides},
  author={Kobayashi, Kaya and Horigane, Kazumasa and Horie, Rie and Akimitsu, Jun},
  booktitle={Physics and Chemistry of Carbon-Based Materials: Basics and Applications},
  pages={149--209},
  year={2019},
  publisher={Springer}
}

@article{simon1997superconductivity,
  title={Superconductivity and chemistry},
  author={Simon, Arndt},
  xjournal={Angewandte Chemie International Edition in English},
  journal={Angew. Chem. Int. Ed.},
  volume={36},
  number={17},
  pages={1788--1806},
  year={1997},
  publisher={Wiley Online Library}
}

@article{schwarz1987band,
  title={Band structure and chemical bonding in transition metal carbides and nitrides},
  author={Schwarz, Karlheinz},
  xjournal={Critical Reviews in Solid State and Material Sciences},
  journal={CRC Crit. Rev. Solid State Mater. Sci.},
  volume={13},
  number={3},
  pages={211--257},
  year={1987},
  publisher={Taylor \& Francis}
}

@article{rotella1983deuterium,
  title={Deuterium site occupation in the oxygen-stabilized $\eta$-carbides \ce{Zr3V3OD_x}. {I}. {P}reparation and neutron powder diffraction},
  author={Rotella, FJ and Flotow, HE and Gruen, DM and Jorgensen, JD},
  xjournal={The Journal of chemical physics},
  journal={J. Chem. Phys.},
  volume={79},
  number={9},
  pages={4522--4531},
  year={1983},
  publisher={American Institute of Physics}
}

@article{RevModPhys.35.1,
  title = {Superconductivity},
  author = {Matthias, B. T. and Geballe, T. H. and Compton, V. B.},
  journal = {Rev. Mod. Phys.},
  volume = {35},
  issue = {1},
  pages = {1--22},
  numpages = {0},
  year = {1963},
  month = {Jan},
  publisher = {American Physical Society},
  doi = {10.1103/RevModPhys.35.1},
  url = {https://link.aps.org/doi/10.1103/RevModPhys.35.1}
}

@article{giorgi1962effect,
  title={Effect of composition on the superconducting transition temperature of tantalum carbide and niobium carbide},
  author={Giorgi, AL and Szklarz, EG and Storms, EK and Bowman, Allen L and Matthias, BT},
  xjournal={Physical Review},
  journal={Phys. Rev.},
  volume={125},
  number={3},
  pages={837},
  year={1962},
  publisher={APS}
}

@article{nowotny1972crystal,
  title={Crystal chemistry of complex carbides and related compounds},
  author={Nowotny, Hans},
  xjournal={Angewandte Chemie International Edition in English},
  journal={Angew. Chem. Int. Ed.},
  volume={11},
  number={10},
  pages={906--915},
  year={1972},
  publisher={Wiley Online Library}
}

@article{weinberger2018review,
  title={Review of phase stability in the group {IVB} and {VB} transition-metal carbides},
  author={Weinberger, Christopher R and Thompson, Gregory B},
  xjournal={Journal of the American Ceramic Society},
  journal={J. Am. Ceram. Soc.},
  volume={101},
  number={10},
  pages={4401--4424},
  year={2018},
  publisher={Wiley Online Library}
}

@inproceedings{stadelmaier1969metal,
  title={Metal-rich metal-metalloid phases},
  author={Stadelmaier, HH},
  booktitle={Developments in the Structural Chemistry of Alloy Phases: Based on a symposium sponsored by the Committee on Alloy Phases of the Institute of Metals Division, the Metallurgical Society, American Institute of Mining, Metallurgical and Petroleum Engineers, Cleveland, Ohio, October, 1967},
  pages={141--180},
  year={1969},
  organization={Springer}
}

@article{shi2024nonmonotonic,
  title={Nonmonotonic superconducting transition temperature and large bulk modulus in the alloy superconductor \ce{Nb4Rh2C_{1-\delta}}},
  author={Shi, Lifen and Ma, Keyuan and Ruan, Binbin and Wang, Ningning and Hou, Jun and Shan, Pengfei and Yang, Pengtao and Sun, Jianping and Chen, Genfu and Ren, Zhian and others},
  xjournal={Physical Review B},
  journal={Phys. Rev. B},
  volume={110},
  number={21},
  pages={214520},
  year={2024},
  publisher={APS}
}

@article{shi2023pressure,
  title={Pressure-driven evolution of upper critical field and {F}ermi surface reconstruction in the strong-coupling superconductor \ce{Ti4Ir2O}},
  author={Shi, Lifen and Ruan, Binbin and Yang, Pengtao and Wang, Ningning and Shan, Pengfei and Liu, Ziyi and Sun, Jianping and Uwatoko, Yoshiya and Chen, Genfu and Ren, Zhian and others},
  xjournal={Physical Review B},
  journal={Phys. Rev. B},
  volume={107},
  number={17},
  pages={174525},
  year={2023},
  publisher={APS}
}

@ARTICLE{shi2025synergetic,
  title     = "Synergetic enhancement of hardness and toughness in new
               superconductors {Ti}$_{2}${Co} and {Ti}$_{4}${Co}$_{2}${O}",
  author    = "Shi, Lifen and Ma, Keyuan and Hou, Jingyu and Ying, Pan and Wang,
               Ningning and Xiang, Xiaojun and Yang, Pengtao and Yu, Xiaohui and
               Gou, Huiyang and Sun, Jianping and Uwatoko, Yoshiya and von Rohr,
               Fabian O and Zhou, Xiangfeng and Wang, Bosen and Cheng, Jinguang",
  journal   = "Chin. Phys. Lett.",
  publisher = "IOP Publishing",
  volume    =  42,
  number    =  6,
  pages     =  067302,
  month     =  jul,
  year      =  2025,
}

@article{yoshimura2025g,
  title={g-Factor Enhanced Upper Critical Field in Superconducting \ce{PdTe2} due to Quantum Confinement},
  author={Yoshimura, Kota and Hsieh, Tzu-Chi and Ma, Huiyang and Chichinadze, Dmitry V and Zou, Shan and Stuckert, Michael and Graf, David and Nowell, Robert and Karim, Muhsin Abdul and Kozawa, Daichi and others},
  journal={arXiv preprint arXiv:2508.07547},
  year={2025}
}

@article{wu2025large,
  title={Large critical fields in superconducting \ce{Ti4Ir2O} from spin-orbit coupling},
  author={Wu, Hao and Shishidou, Tatsuya and Weinert, Michael and Agterberg, Daniel F},
  xjournal={Physical Review B},
  journal={Phys. Rev. B},
  volume={111},
  number={18},
  pages={184506},
  year={2025},
  publisher={APS}
}

@article{norman1990magnetic,
  title={Magnetic quantization and the upper critical field of superconductors},
  author={Norman, MR},
  xjournal={Physical Review B},
  journal={Phys. Rev. B},
  volume={42},
  number={10},
  pages={6762},
  year={1990},
  publisher={APS}
}

@article{altarawneh2012superconducting,
  title={Superconducting pairs with extreme uniaxial anisotropy in \ce{URu2Si2}},
  author={Altarawneh, MM and Harrison, N and Li, G and Balicas, L and Tobash, PH and Ronning, F and Bauer, ED},
  xjournal={Physical review letters},
  journal={Phys. Rev. Lett.},
  volume={108},
  number={6},
  pages={066407},
  year={2012},
  publisher={APS}
}

@article{Fang2012,
  title = {Stability and crystal structures of iron carbides: A comparison between the semi-empirical modified embedded atom method and quantum-mechanical {DFT} calculations},
  author = {Fang, C. M. and van Huis, M. A. and Thijsse, B. J. and Zandbergen, H. W.},
  journal = {Phys. Rev. B},
  volume = {85},
  issue = {5},
  pages = {054116},
  numpages = {7},
  year = {2012},
  month = {Feb},
  publisher = {American Physical Society},
  doi = {10.1103/PhysRevB.85.054116},
  url = {https://link.aps.org/doi/10.1103/PhysRevB.85.054116}
}

@Article{Yu2024,
author={Yu, Jieru
and Du, Jinglian
and Shang, Shun-Li
and Fu, Hejian
and Hao, Yang
and He, Liubaixiang
and Liu, Zi-Kui
and Liu, Feng},
title={Tailoring the stability of iron carbides to enhance the mechanical performances of {F}e--{C}--{M}n--{S}i alloys},
xjournal={Journal of Materials Science},
journal={J. Mater. Sci.},
year={2024},
month={Jun},
day={01},
volume={59},
number={24},
pages={11157-11176},
issn={1573-4803},
doi={10.1007/s10853-024-09824-w},
xurl={https://doi.org/10.1007/s10853-024-09824-w}
}

@book{rudman1967phase,
  title={Phase stability in metals and alloys},
  author={Rudman, Peter S and Stringer, John and Jaffee, Robert Isaac},
  volume={1},
  year={1967},
  publisher={McGraw-Hill}
}

@book{toth2014transition,
  title={Transition metal carbides and nitrides},
  author={Toth, Louis},
  year={2014},
  publisher={Elsevier}
}

@article{Shi2026Ti4Co2O_pressure,
  title={Two distinct superconducting regimes in Ti4Co2O under pressures},
  author={Shi, Lifen and Ma, Keyuan and Ruan, Binbin and Wang, Zhen and Yang, Pengtao and Ren, Zhian and Sun, Jianping and Li, Gang and von Rohr, Fabian O and Wang, Bosen and others},
  journal={arXiv preprint arXiv:2605.01893},
  year={2026}
}

@article{Koepernik1999,
  author  = {Koepernik, Klaus and Eschrig, Helmut},
  title   = {{Full-potential nonorthogonal local-orbital minimum-basis band-structure scheme}},
  journal = {Phys. Rev. B},
  volume  = {59},
  pages   = {1743--1757},
  year    = {1999},
  doi     = {10.1103/PhysRevB.59.1743}
}

@article{Perdew1996,
  author  = {Perdew, John P. and Burke, Kieron and Ernzerhof, Matthias},
  title   = {{Generalized gradient approximation made simple}},
  journal = {Phys. Rev. Lett.},
  volume  = {77},
  pages   = {3865--3868},
  year    = {1996},
  doi     = {10.1103/PhysRevLett.77.3865}
}

@article{Daeves1921,
author = {Daeves, Karl},
title = {Grenzen der {L}öslichkeit für {K}ohlenstoff in ternären {S}tählen. {II}. {D}as {S}ystem {C}hrom-{E}isen-{K}ohlenstoff},
xjournal = {Zeitschrift für anorganische und allgemeine Chemie},
  journal   = "Z. Allg. Anorg. Chem.",
volume = {118},
number = {1},
pages = {67-74},
doi = {10.1002/zaac.19211180105},
xurl = {https://onlinelibrary.wiley.com/doi/abs/10.1002/zaac.19211180105},
xeprint = {https://onlinelibrary.wiley.com/doi/pdf/10.1002/zaac.19211180105},
year = {1921}
}

@article{Takeda1930a,
  title={A Metallographic Investigation of the Ternary Alloys of the Iron-Tungsten-Carbon System. {I}. {O}n the Carbides in Tungsten Steels},
  author={Takeda, Shuz{\^o}},
  xjournal={Tech. Report of Tohuku Imperial University},
  journal={Tech. Rept. Tohuku Imp. Univ.},
  volume={9},
  pages={483-514},
  year={1930}
}

@article{Takeda1930b,
  title={A Metallographic Investigation of the Ternary Alloys of the Iron-Tungsten-Carbon System. {II}. {T}ransformation and Constitution of Tungsten Steels},
  author={Takeda, Shuz{\^o}},
  xjournal={Tech. Report of Tohuku Imperial University},
  journal={Tech. Rept. Tohuku Imp. Univ.},
  volume={9},
  pages={627-664},
  year={1930}
}

@article{Takeda1931,
  title={A Metallographic Investigation of the Ternary Alloys of the Iron-Tungsten-Carbon System. {III}. {T}he Equilibrium Diagram of the \ce{Fe-W-C} System},
  author={Takeda, Shuz{\^o}},
  xjournal={Tech. Report of Tohuku Imperial University},
  journal={Tech. Rept. Tohuku Imp. Univ.},
  volume={10},
  pages={42-92},
  year={1931}
}

@article{takeda1936metallographic,
  title={A metallographic study of the action of the cementing materials for cemented tungsten carbide},
  author={Takeda, Shuz{\^o}},
  journal={Science Rept Tohoku Univ, Honda Anniv},
  pages={864--881},
  year={1936}
}

@article{Adelskoeld1933,
author = {Adelsköld, V. and Sundelin, A. and Westgren, A.},
title = {Carbide in kohlenstoffhaltigen Legierungen von Wolfram und Molybdän mit Chrom, Mangan, Eisen, Kobalt und Nickel},
journal = {Zeitschrift für anorganische und allgemeine Chemie},
volume = {212},
number = {4},
pages = {401-409},
year = {1933}
}

\end{document}